\documentclass[iop]{emulateapj}

\usepackage{graphicx}
\usepackage{color}
\usepackage{url}
\usepackage{hyperref}
\usepackage{enumitem}
\usepackage{natbib}
\usepackage{mathrsfs}
\usepackage{mathpazo}
\usepackage{bm}
\usepackage{amsmath}

\newcommand{\beq}{\begin{equation}}
\newcommand{\eeq}{\end{equation}}
\newcommand{\beqn}{\begin{eqnarray}}
\newcommand{\eeqn}{\end{eqnarray}}

\newcommand{\ee}{\mathrm{e}}
\newcommand{\Ma}{\mathrm{Ma}}
\newcommand{\Rey}{\mathrm{Re}}
\newcommand{\Kn}{\mathrm{Kn}}

\newcommand{\xtimes}[2]{#1\times{10^{#2}}}

\newcommand{\vt}[1]{\boldsymbol{\mathrm{#1}}}       
\renewcommand{\v}[1]{{\boldsymbol{#1}}} 

\renewcommand{\sun}{\odot}

\newcommand{\hatr}{\hat{\v{r}}}

\newcommand{\hatphi}{\hat{\v{\phi}}}
\newcommand{\hatz}{\hat{\v{z}}}
\newcommand{\haty}{\hat{\v{y}}}
\newcommand{\hatx}{\hat{\v{x}}}

\newcommand{\degree}{\ensuremath{^\circ}}

\newcommand{\Eq}[1]{Eq.~(\ref{#1})}

\newcommand{\Eqss}[2]{Eqs.~(\ref{#1})--(\ref{#2})}
\newcommand{\eq}[1]{\Eq{#1}}
\newcommand{\eqp}[1]{(Eq.~\ref{#1})}

\newcommand{\eqss}[2]{\Eqss{#1}{#2}}

\newcommand{\Fig}[1]{Fig.~\ref{#1}}
\newcommand{\fig}[1]{\Fig{#1}}

\newcommand{\sect}[1]{Sect.~\ref{#1}}

\def\apj{\rm ApJ}
\def\apjl{\rm ApJL}

\def\apss{\rm ApSS}
\def\aj{\rm AJ}
\def\mnras{\rm MNRAS}
\def\nat{\rm Nature}

\def\aap{\rm A\&A}
\def\araa{\rm ARA\&A}
\def\icarus{\rm Icarus}

\def\planss{\rm Planet. Space Sci.}

\newcommand{\hatq}{\hat{\v{q}}}
\newcommand{\hath}{\hat{\v{h}}}
\newcommand{\hate}{\hat{\v{e}}}

\newcommand{\mean}[1]{\left\langle #1 \right\rangle}
\newcommand{\sinf}{\sin{f}}
\newcommand{\cosf}{\cos{f}}
\newcommand{\sinfw}{\sin{\left(f-\omega\right)}}
\newcommand{\cosfw}{\cos{\left(f-\omega\right)}}
\newcommand{\cosE}{\cos{E}}
\newcommand{\teff}{\tau_{\rm eff}}

\newcommand{\sinI} {\sin{I}}
\newcommand{\cosI} {\cos{I}}
\newcommand{\cosO}{\cos{\Omega}}
\newcommand{\sinO}{\sin{\Omega}}
\newcommand{\sinw}{\sin{\omega}}
\newcommand{\cosw}{\cos{\omega}}

\newcommand{\sinisq}{\sin^2{I}}
\newcommand{\sintwow}{\sin{2\omega}}
\newcommand{\costwow}{\cos{2\omega}}
\newcommand{\sintwop}{\sin{2\Psi}}
\newcommand{\costwop}{\cos{2\Psi}}

\shorttitle{Orbital evolution of MU69}
\shortauthors{Lyra et al.}

\begin{document}

\title{Evolution of MU69 from a binary planetesimal into contact\\ 
by Kozai-Lidov oscillations and nebular drag}

\author{Wladimir Lyra\altaffilmark{1}, Andrew
  N. Youdin\altaffilmark{2}, Anders Johansen\altaffilmark{3}}
\altaffiltext{1}{Department of Astronomy, New Mexico State University, PO BOX 30001, MSC 4500, Las Cruces, NM 88003-8001.}
\altaffiltext{2}{Steward Observatory, University of Arizona, Tucson, AZ, 85721}
\altaffiltext{3}{Lund Observatory, Department of Astronomy and Theoretical Physics, Lund University, Box 43, 221 00 Lund, Sweden}

\begin{abstract}
The New Horizons flyby of the cold classical Kuiper Belt 
object MU69 showed it to be a contact binary. The existence 
of other contact binaries in the 1--10\,km range raises the question of how 
common these bodies are and how they evolved into contact. Here we
consider that the pre-contact lobes of MU69 formed as a binary embedded in the
Solar nebula, and calculate its subsequent orbital evolution in the
presence of gas drag.  We find that the sub-Keplerian wind of the disk brings the drag timescales for
10\,km bodies to under 1\,Myr for quadratic-velocity drag, which is valid in the
asteroid belt. In the Kuiper belt, however, the drag is linear
  with velocity and the effect of the wind cancels out as the angular
  momentum gained in half an orbit is exactly lost in the other half; the drag
  timescales for 10\,km bodies remain $\gtrsim$ 10 Myr. In
this situation we find that a combination of nebular drag and Kozai-Lidov
oscillations is a promising channel for collapse. We analytically solve the hierarchical three-body problem with nebular drag and
implement it into a Kozai cycles plus tidal friction model. The
permanent quadrupoles of the pre-merger lobes make the Kozai oscillations
stochastic, and we find that when gas drag is included the shrinking of the
semimajor axis more easily allows the stochastic fluctuations to bring
the system into contact. Evolution to contact happens very rapidly
(within $10^4$\,yr) in the pure, double-average quadrupole,
Kozai region between $ \approx 85-95\degree$, and within 3\,Myr in the
drag-assisted region beyond it. The synergy between
$J_2$ and gas drag widens the window of contact to
80$\degree$--100$\degree$ initial inclination, over a larger range of semimajor axes than Kozai and 
$J_2$ alone. As such, the model predicts a low initial occurrence of binaries
in the asteroid belt, and an initial contact binary fraction of about
10\% for the cold classicals in the Kuiper belt. The speed at contact is
the orbital velocity; if contact happens at pericenter at high
eccentricity, it deviates from the escape velocity only because of the
oblateness, independently of the semimajor axis. For MU69, the oblateness leads to a 30\% decrease in
contact velocity with respect to the escape velocity, the latter scaling with the
square root of the density. For mean densities in the range
0.3-0.5 g\,cm$^{-3}$, the contact velocity should be $3.3-4.2$
m\,s$^{-1}$, in line with the observational
evidence from the lack of deformation features and estimate of the
tensile strength. 
\end{abstract}

\section{Introduction}

On Jan 1st 2019 the New Horizons spacecraft flew past 2014 MU69 (hereafter referred to as MU69), a small
($\approx$ 30\,km) trans-Neptunian object, recently renamed ``Arrokoth''. Its low-eccentricity and low-inclination
orbit identifies it as a ``cold classical'' Kuiper Belt object \cite[CCKBO,][]{Brown01,Kavelaars+08,Petit+11}. Unlike the heavily processed
comet 67P/Churyumov-Gerasimenko visited by the Rosetta mission, MU69 is presumably a pristine planetesimal
kept undisturbed for the entirety of its 4.6 Gyr residence in the Kuiper belt. 

The flyby showed MU69 to be a contact binary where the two lobes have dimensions $20.6\times 19.9 \times 9.4$\,km and
$15.4\times 13.8\times 9.8$\,km \citep[$\pm$0.5$\times$0.5$\times$2,][]{Stern+19}. Their similar
colors and composition, as well as axial alignment 
indicate that the individual lobes formed close to one another, and underwent
orbital evolution that led to contact. The close formation is backed
by observational data suggesting a high binary fraction among CCKBOs \citep[30\%, and possibly larger due to
observational limitation,][]{Noll+08a,Veillet+02,Petit+08,Grundy+11,Fraser+17}. Nearly
equal-sized contact binaries represent 10\%-25\% of cold
classicals \citep{ThirouinSheppard19}. Given the lack of major
deformations and estimates of the tensile strength \citep{JutziAsphaug15,McKinnon+19,Wandel+19}, the contact must have happened at
low speeds, below the escape velocity ($\lesssim$6\,m/s). 

The formation of the individual lobes could be the result of a gravitational
instability of solids in the disk midplane
\citep{GoldreichWard73,YoudinShu02}, and indeed \cite{Nesvorny+10} showed that
gravitational collapse can produce binaries with order unity mass
ratios.  The collapse model predicts that the composition and colors
of binary partners should match, which is confirmed by observations of
(non-contact) Kuiper Belt binaries \citep{Benecchi+09} as well as MU69.
More specifically, gravitationally collapse can be seeded by the
streaming instability \citep{YoudinGoodman05,Johansen+07,YoudinJohansen07,JohansenYoudin07},
which has been recently shown to lead preferentially to binary
planetesimals, also matching the ratio of prograde to
retrograde mutual inclination among Kuiper belt binaries \citep{Nesvorny+19}. In this paper we consider the lobes already formed, 
and examine the subsequent orbital evolution.

Immediately after formation as a binary, if the system has a
high enough inclination with respect to the ecliptic, Kozai-Lidov oscillations
\citep{Lidov62,Kozai62} could lead to binary
coalescence \citep{MazehShaham79,Nesvorny+03,PeretsNaoz09,Naoz+10}. The Kozai-Lidov effect is a well-studied resonance occurring in triple
systems \citep[for a recent review, see][]{Naoz16}, whereby
eccentricity and  inclination undergo periodic oscillations. The system is
considered hierarchical in scale if the triple system is composed of two
binaries with clear separation of scales, with two of the bodies composing a
tight inner binary (semimajor axis $a$), and the inner binary and third body composing a
wide outer binary (semimajor axis $a_{\rm out}\gg a$). In the case of MU69, the two pre-merger lobes are
the inner binary ($a$ presumably of the order of $10^3$-$10^4$\,km), and their center of mass orbiting the distant Sun is
the outer binary ($a_{\rm out} =45$\,AU). The angular momentum of the system is the sum of the
angular momenta of the inner ($\v{h}$) and outer ($\v{h}_{\rm out}$)
binaries. Considering the vertical $z$ direction to be along the
direction of total angular momentum $\v{H} = \v{h}+\v{h}_{\rm out}$,
then conservation of $\v{H}$ implies that the $z$-projected
angular momentum $(h_z+h_{{\rm out},z})$ is conserved. If we make
the further approximation that $\v{h}_{\rm out}$ is conserved, then $h_z$ must
be conserved as well. The $z$-component is proportional to $H_k \equiv \cos I \sqrt{1-e^2}$,
dubbed Kozai constant, where $e$ is the eccentricity of the inner
binary and $I$ is its inclination with respect to the outer binary; because $H_k$ is
constant but not $e$ or $I$, there exists the possibility of
exchanging $I$ for $e$ and vice-versa. 

For highly inclined orbits, the Kozai-Lidov resonance can drive very
high eccentricity. This mechanism has been invoked to explain why no irregular satellites
have inclinations in the range 50$\degree$-140$\degree$
\citep{Nesvorny+03}: such satellites would be driven by Kozai cycles
either to pericenters that impact the planets (or massive inner
moons), or to apocenters that lie outside the Hill sphere
\citep{Carruba+02}. \cite{ThomasMorbidelli96}, applying it to a
triple system of Sun-Jupiter-comet, showed that Kozai-Lidov
oscillations can make cometary orbits become Sun-grazing.

All of the above assumes that the bodies are point masses. Yet tidal
friction cannot be ignored for $\approx$10\,km objects, not only because
these bodies are deformable, but also because their significant deviations from spherical
symmetry mean that they have permanent quadrupoles. As the orbiters approach each
other, tidal friction should drive circularization and orbital decay
for retrograde orbiters. However, during Kozai cycles the longitude of
pericenter $\omega$ librates about either 90$\degree$ or
270$\degree$ \citep{Naoz16}; if the magnitude $J_2$ of the quadrupole potential is strong enough, it will cause
precession and unlock $\omega$ from the libration, frustrating the
resonance. \cite{PorterGrundy12} conclude that, in the presence of
tides, Kozai cycles can collapse high inclination binaries only if the
semimajor axis is above a critical value that depends on the strength
of $J_2$. This value is placed at the critical semimajor axis at
$a_{\rm crit} > 0.05 R_H$, where $R_H$ is the radius of the Hill sphere, based on the
observation that many known Kuiper Belt binaries with $a < 0.05 R_H$ have
mutual inclination around 90$\degree$. 

In this work we are concerned with the effect of nebular drag on a freshly
formed binary planetesimal. We ask if nebular drag can by itself
collapse a binary or, in the negative, if it can affect the Kozai cycles that would or would not lead to
contact in the absence of gas. If formation was triggered by the streaming
instability as models suggest \citep{Nesvorny+19}, then MU69 formed while the
Solar Nebula was still present, and nebular drag should have impacted
its orbital evolution. 

Evidence that MU69 formed in the presence of
gas is shown by \cite{Lisse+19}, studying the stability of ices in its
surface. The ices in spectrum of MU69 show presence of water ice,
methanol, and HCN. Pluto, on the other hand, shows CH$_4$, N$_2$, and
and CO, which were searched for in MU69 and not
found. \cite{Lisse+19} explored the thermodynamics of laboratory
ices, showing that at the temperature of MU69 \citep[with a night-day range
of 16K-58K, and average 35K,][]{Umurhan+19} the near-vacuum sublimation
rate of the main volatiles is such that only highly refractory,
hydrogen bonded species such as water and methanol
survive for 4.5\,Gyr. The hypervolatiles CH$_4$, N$_2$, and
and CO should be lost in under 1\,Myr and should never have
been incorporated into small KBOs unless the temperature at formation
was much colder than the present equilibrium temperature (but see \citealt{Krijt+18}). 
For N$_2$ in particular, the temperature must have been  $\approx$
15\,K.  Their presence on the surface of Pluto is due to gravitational
retention; yet, if Pluto was formed out of millions of MU69-like bodies, then
these bodies must have had these hypervolatiles. The contradiction can
be resolved if MU69 was formed in an environment of much lower
temperatures, as it should be expected if it was
formed in the optically thick confines of a protoplanetary disk.

We therefore explore the orbital evolution of the pre-merger lobes under nebular drag, mutual gravitational
interaction, and solar tides. We implement a Kozai
cycle plus tidal friction model, following the formalism of
\citet[][see also
\citealt{FabryckyTremaine07,PeretsFabrycky09,PorterGrundy12}]{EggletonKiseleva01}. 
In addition to these processes we add the permanent $J_2$ quadrupole as
derived by \cite{Ragozzine09} and our implementation of nebular drag
from the Solar Nebula, that we derive in this work. We call the full model KTJD, for Kozai cycles plus tidal friction plus $J_2$ plus drag. 

This paper is structured as follows. In \sect{sect:model} we describe
the model, in \sect{sect:results} the results. We conclude in
\sect{sect:conclusion} with a discussion and summary of these
results. Involved mathematical details are shown in appendices.

\section{Model}
\label{sect:model}

\begin{figure}
  \begin{center}
    \resizebox{\columnwidth}{!}{\includegraphics{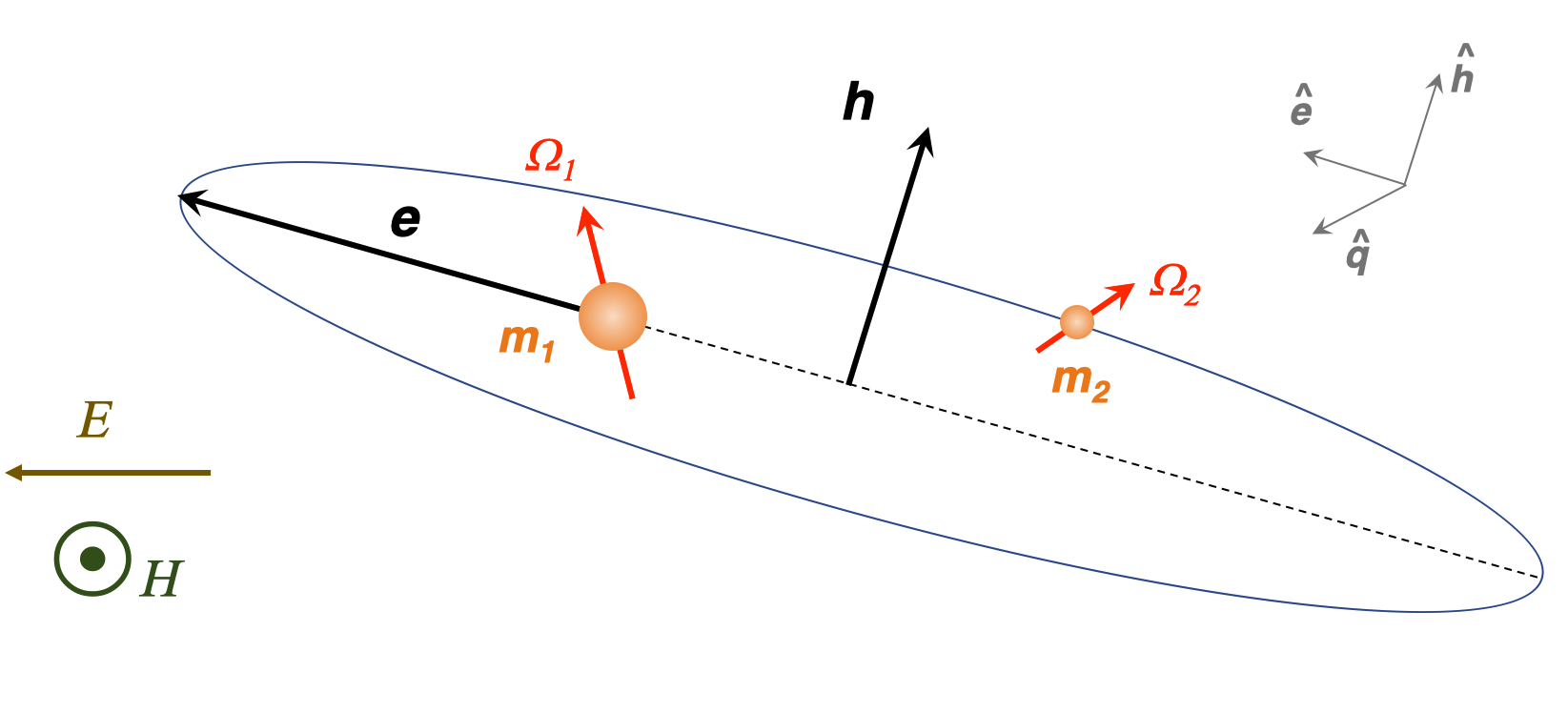}}
  \end{center}
\caption{Orientation of the vectors for the Kozai cycles plus tidal
  friction model. The model solves for the eccentricity vector $\v{e}$, the
  angular momentum $\v{h}$ of the inner binary, and the spin angular
  momentum of each body, $\v{\Omega}_1$ and $\v{\Omega}_2$. The vectors
  $\v{e}$, $\v{h}$, and $\v{q}=\v{h}\times\v{e}$ define a system of
  time-varying orthogonal bases. The eccentricity and angular momentum
vectors of the outer orbit, $\v{E}$, and $\v{H}$, remain constant. The
orbital inclination is the angle between $\v{h}$ and $\v{H}$.}
\label{fig:kozaigeometry}
\end{figure}

\begin{figure}
  \begin{center}
    \resizebox{\columnwidth}{!}{\includegraphics{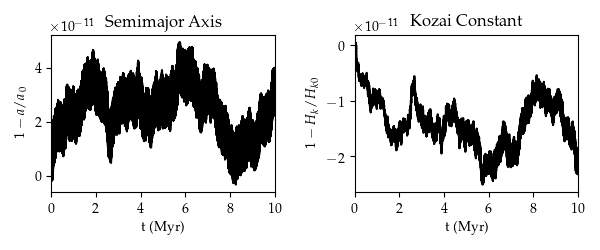}}
\end{center}
\caption{Conservation of the semimajor axis and of the Kozai constant
  $H_k = \cos{I} \sqrt{1-e^2}$ related to the vertical angular
  momentum. With an adaptative timestep responding to period and
  eccentricity, we achieve conservation down to $10^{-11}$ over 10\,Myr.}
\label{fig:kozai_conservation}
\end{figure}

In this work we use two different codes. The first is a $N$-body model that
solves for position and velocities of point masses under mutual gravitational
interaction. The second is a Kozai cycle plus tidal friction model, which evolves the
eccentricity vector and the orbital angular momentum of the
binary, along with the spin angular momenta of the two bodies, while
keeping the external, heliocentric, orbit constant (see
\fig{fig:kozaigeometry}). Both use a standard 3rd order 
Runge-Kutta, i.e., the accumulated error is proportional to the
  cube of the timestep. 

\subsection{The KTJD model}

We follow the equations of \cite{EggletonKiseleva01} in the reference frame
of the orbit of the binary. The (time-varying) orientation is given by the unit
vectors $\hate$ pointing to pericenter, $\hath$ pointing to the direction of orbital
angular momentum, and $\hatq=\hath \times \hate$ along the latus
rectum (see \fig{fig:kozaigeometry}). 

We split the model equations for the eccentricity and angular momentum
vectors into equations for their moduli and unit vectors,
respectively. The full model including the evolution of the spin
vectors consists of 14 coupled equations 

\beqn
\frac{de}{dt} &=& - e\left[V_1 + V_2  + V_d  + 5\left(1-e^2\right) S_{eq}\right],\label{eq:model1}\\
\frac{dh}{dt} &=& - h\left(W_1 + W_2 + W_d - 5e^2S_{eq}\right),\\
\frac{d\hate}{dt} &=& \left[Z_1 + Z_2 + \left(1-e^2\right)\left(4S_{ee}-S_{qq}\right)\right]\hatq \nonumber\\
&-&\left[Y_1 + Y_2 + \left(1-e^2\right)S_{qh}\right]\hath,\\
\frac{d\hath}{dt} &=& \left[Y_1 + Y_2 + \left(1-e^2\right)S_{qh}\right]\hate
\nonumber\\
&-&\left[X_1 + X_2 +          \left(4e^2+1\right)S_{eh}\right]\hatq,\label{eq:model4}
\eeqn
\beqn
\frac{d\v{\Omega}_1}{dt} &=& \frac{\mu_r h}{\mathcal{I}_1} \left(-Y_1\hate + X_1\hatq + W_1\hath\right), \\
\frac{d\v{\Omega}_2}{dt} &=& \frac{\mu_r h}{\mathcal{I}_2} \left(-Y_2\hate + X_2\hatq + W_2\hath\right).
\eeqn

\noindent Here the indices 1 and 2 refer to each orbiter of the
inner binary. The quantities $V_i$ and $W_i$ ($i=1,2$) are dissipative
functions related to how a deformable body responds to a tidal
field. The quantities $X$, $Y$, $Z$ give precession and apsidal motion. The tensor $S_{ij}$ relates to the 3rd body and is responsible for the Kozai cycles, here added up to the
quadrupole level of approximation \citep{Kiseleva+98} and keeping the
outer orbit exactly constant ($a_{\rm out}=45$\,AU and $e_{\rm out}=0.04$ for MU69). In the model, the outer orbit is specified by
the time-independent vectors $\v{H}$ and $\v{E}$, the angular momentum
and eccentricity vectors of the outer orbit, respectively. 

We refer the reader to
\cite{EggletonKiseleva01} for the detailed mathematical form of the
parameters $X$, $Y$, $Z$, $V$, $W$, and $S_{ij}$, and to
\cite{Ragozzine09}  to how the planetary permanent quadrupole 
impacts these parameters, depending also on the rigidity $\mu_{\rm b}$ of the body and the tidal dissipation
quality factor $Q$. The parameter $\mathcal{I}_i= 0.4 m_i(R_{xi}^2+R_{yi}^2)$
is the moment of inertia of body $i$ about its spin axis, 
and we do not consider non-principal axis rotators; $m_i$ is the
mass of body $i$; $R_{xi}$ and $R_{yi}$ are the
principal semiaxes of body $i$ perpendicular to the spin
axis. Finally, $\mu_r=m_1m_2/(m_1+m_2)$ is the reduced mass.

The integration timestep is adaptive with semimajor axis and eccentricity, 

\beq
\Delta t = C T \sqrt{\frac{1-e}{1+e}}
\eeq

\noindent where $T$ is the orbital period of the inner binary, updated
as it hardens. The factor in the square root is the velocity at pericenter normalized by
the circular velocity. We find this important to conserve energy and
angular momentum during the high-eccentricity excursions. We test a
non-dissipative, purely Kozai cycle model, to 10\,Myr, and find that
with $C=10^{-2}$ the semimajor axis and the Kozai constant are conserved and bounded to one part in
$10^{11}$ (\fig{fig:kozai_conservation}). We test that decreasing $C$
to $10^{-3}$ does not improve conservation, and for $C=10^{-1}$ the error
grows. We use thus $C=10^{-2}$ in all integrations. The code,
  written in Fortran90, is made
public and can be downloaded from \url{https://github.com/wlyra/yoshikozai} 

\begin{figure}
  \begin{center}
    \resizebox{\columnwidth}{!}{\includegraphics{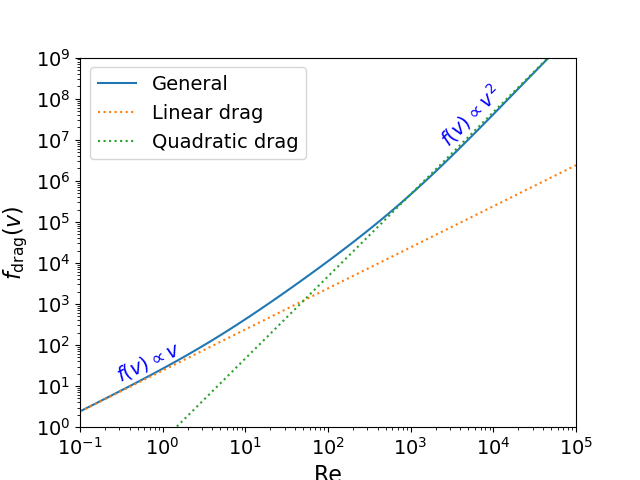}}
\end{center}
\caption{Dependency of the drag regime ($F_{\rm drag} \propto C_D \Rey^2$) on Reynolds number. The drag is linear in the
  viscous regime (Stokes law) and quadratic in the turbulent (ram
  pressure) regime, with a smooth transition in between. MU69 in the
  MMSN lies at $\Rey\approx 10$, closer to the linear range.}
\label{fig:drag}
\end{figure}

\begin{figure*}
  \begin{center}
    \resizebox{.475\textwidth}{!}{\includegraphics{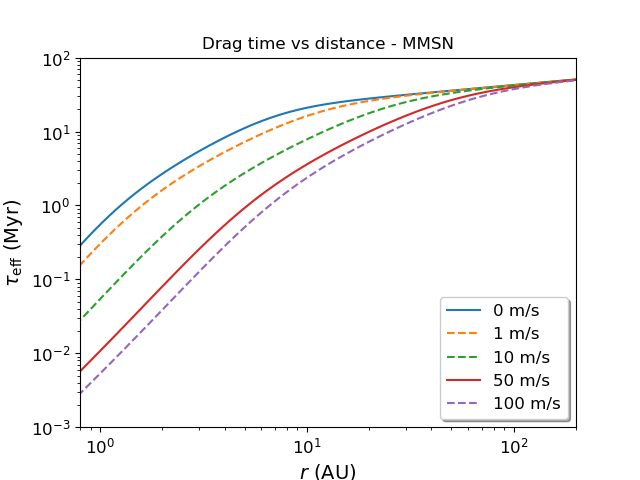}}
    \resizebox{.475\textwidth}{!}{\includegraphics{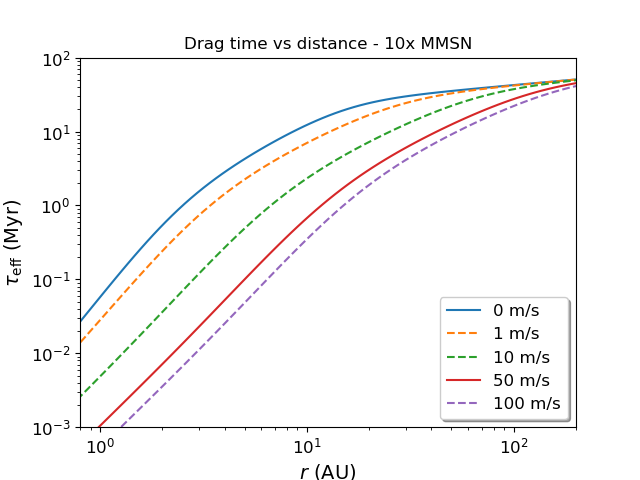}}
    \resizebox{.475\textwidth}{!}{\includegraphics{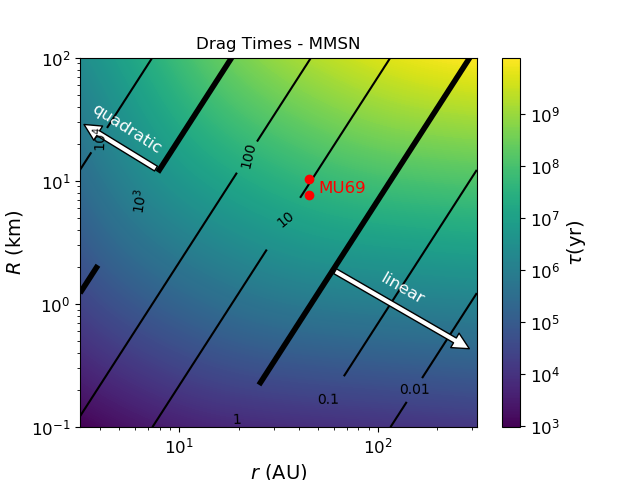}}
    \resizebox{.475\textwidth}{!}{\includegraphics{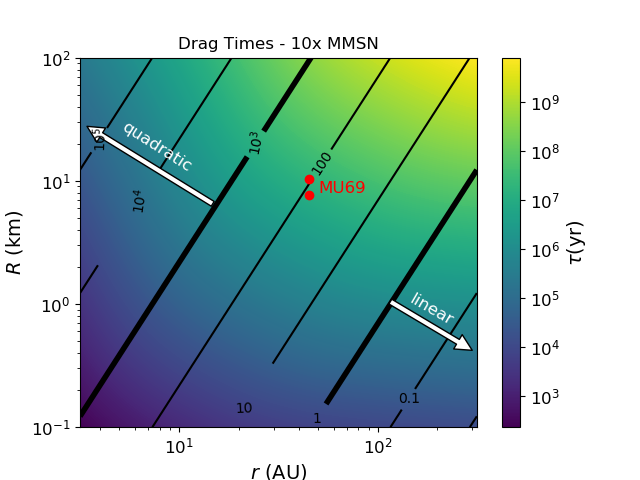}}
\end{center}
\caption{Upper panels: Effective drag time vs distance for bodies the
  size and mass of MU69 as function of the wind velocity in the MMSN
  (left) and a solar nebula ten times more massive (right). The
  red curve (50 m/s) corresponds to standard pressure gradients, zero
  (blue) to a local pressure maximum, and others speeds to reduced or
  enhanced pressure gradients. For the
  no-wind curve (blue) only the binary velocity is considered.
  Stokes drag, where the wind has no effect, is valid in the Kuiper
  belt. In the asteroid belt the Reynolds number place the drag force
  in the quadratic regime, and the wind brings the effective drag time down to
  0.1\,Myr. Nebular drag can collapse an object like MU69 in the asteroid
  belt in the lifetime of the Solar Nebula, but not in the Kuiper
  belt. Lower panels: Drag times as a function of distance (x-axis)
    and body radius (y-axis) in the MMSN (left) and a solar nebular
    ten times more massive (right), assuming a 50 m/s wind. The color maps refers to drag
    times. The lines to Reynolds numbers. MU69 lies in the transition
    between linear drag (${\rm Re} < 1$) and quadratic drag (${\rm Re} > 1000$).}
\label{fig:drag_linearvsquadratic}
\end{figure*}

\begin{figure*}
  \begin{center}
    \resizebox{\textwidth}{!}{\includegraphics{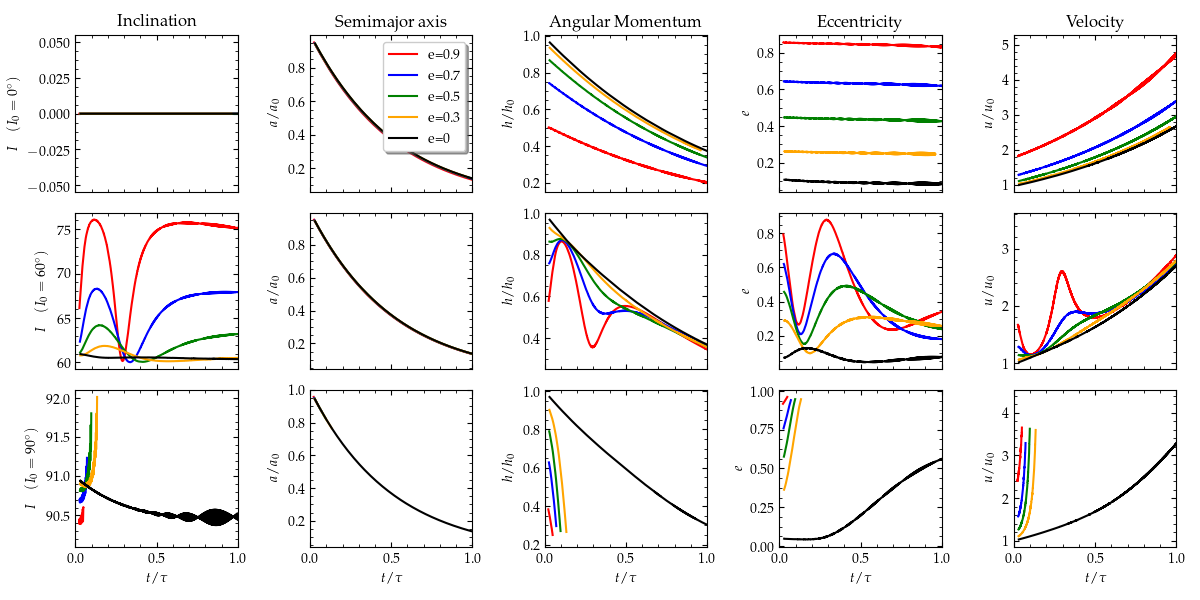}}
\end{center}
\caption{$N$-body evolution of a binary of initial inclination
  $0\degree$ (top panels), $60\degree$ (middle panels) and
  $90\degree$ (bottom panels) for a range of initial
  eccentricities. The lines are box-averaged over a solar period. Time
  is normalized by the friction time $\tau$. The long-term evolution in zero inclination
  is well described by the analytical solution for
  $a$,$e$, $h$ averaged over a solar orbit. The system comes to
  contact within the timescale set by $\tau$. This is appropriate for
    contact within the lifetime of the Solar nebula for the range of semimajor axes of the asteroid belt ($\tau_{\rm eff} \approx 0.1$\,Myr), but would not lead to contact in
  the Kuiper Belt ($\tau_{\rm eff} > 10$\,Myr). For $I_0=60\degree$ one
  Kozai cycle is seen to occur. Yet, the eccentricity does not rise
  high enough to lead to contact. For $I_0=90\degree$ initial
  inclination rapid evolution to contact
  happens, even for moderately low eccentricities ($e\gtrsim0.3$). The
  evolution is of very fast fall of angular momentum and increase of
  eccentricity with nearly constant semimajor axis, plunging the
  binary into contact through a nearly radial trajectory.}
\label{fig:incl00}
\end{figure*}

\begin{figure}
  \begin{center}
    \resizebox{\columnwidth}{!}{\includegraphics{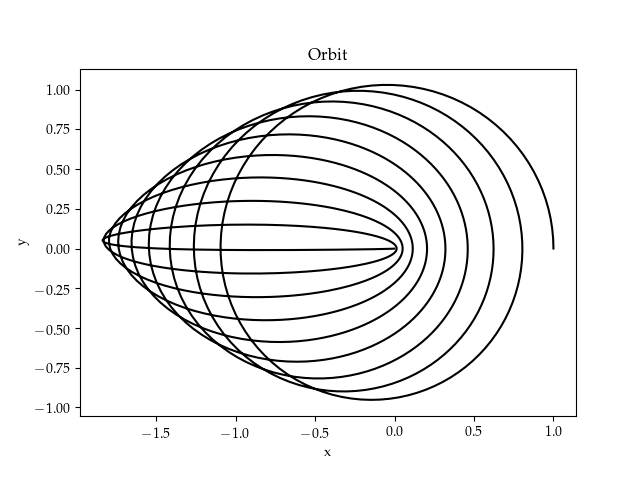}}
\end{center}
\caption{The trajectory to contact in the reference frame of the
  primary of the inner binary, oriented along the $\hate\hatq\hath$ vectors, here fixed. The secondary started
  at ($x,y$)=(1,0) and ended at the origin. Its trajectory was one of
  flattening the $y$ axis while the magnitude of the semimajor axis
  remains roughly constant, increasing the eccentricity and losing
  angular momentum.}
\label{fig:trajectory}
\end{figure}

\begin{table*}
\caption[]{Symbols used in this work.}
\label{table:symbols}
\begin{center}
\begin{tabular}{lll c lll}\hline
Symbol                      & Definition                                                                & Description                              && Symbol                    & Definition                                              & Description\\\hline
\small{$a$}                  &\small{$$}
                                                                                                        &\small{semimajor
                                                                                                          axis
                                                                                                          of
                                                                                                          inner
                                                                                                          binary}
                                                                                                                                                   &&\small{$\Delta\v{v}$}     &\small{$ \v{v}-\v{u} $}                                   &\small{velocity of flow past object}\\
\small{$a_{\rm out}$}        &\small{$$}                                                                  &\small{semimajor axis of outer binary}     &&\small{$d\v{F}$}            &\small{$$}                                                &\small{force differential}\\                                        
\small{$h$}                  &\small{$$}                                                                  &\small{inner binary angular momentum}      &&\small{$\hatr$}             &\small{$$}                                                &\small{cylindrical rotation of $\hate$}\\           
\small{$h_{\rm out}$} &\small{$$}                                                                  &\small{outer binary angular momentum}      &&\small{$\hatphi$}           &\small{$$}                                                &\small{cylindrical rotation of $\hatq$}\\               
\small{$H$}                  &\small{$h+h_{\rm out}$}                                              &\small{total angular momentum}             &&\small{$\overline{R}$}      &\small{$d\v{F}\cdot\hatr$}                                &\small{radial force}\\                                          
\small{$t$}                  &\small{$$}                                                                  &\small{time}                               &&\small{$\overline{T}$}      &\small{$d\v{F}\cdot\hatphi$}                              &\small{azimuthal force}\\                                       
\small{$I$}                  &\small{$\cos^{-1}(h_z/h)$}                                                  &\small{mutual inclination of inner binary} &&\small{$\overline{N}$}      &\small{$d\v{F}\cdot\hath$}                                &\small{vertical force}\\                                        
\small{$e$}                  &\small{$$}                                                                  &\small{eccentricity of inner binary}       &&\small{$f$}                 &\small{$$}                                                &\small{true anomaly}\\                                          
\small{$H_k$}                &\small{$\cos I \sqrt{1-e^2}$}                                               &\small{Kozai constant}                     &&\small{$E$}                 &\small{$M=E-e\sin E$}                                     &\small{eccentric anomaly}\\                                          
\small{$\omega$}             &\small{$$}                                                                  &\small{longitude of pericenter}            &&\small{$M$}                 &\small{$nt$}                                              &\small{mean anomaly}\\                                               
\small{$m_1$}                &\small{$$}                                                                  &\small{mass of primary}                    &&\small{$\vt{R}$}            &\small{$$}                                                &\small{Rotation matrix}\\                                            
\small{$m_2$}                &\small{$$}                                                                  &\small{mass of secondary}                  &&\small{$m_b$}               &\small{$m_1 + m_2$}                                       &\small{sum of masses of bodies}\\                               
\small{$M_\sun$}             &\small{$$}                                                                  &\small{solar mass}                         &&\small{$R_b$}               &\small{$R_1 + R_2$}                                       &\small{sum of radii of bodies}\\                                
\small{$R_H$}                &\small{$a \left(\frac{m_1+m_2}{3M_\sun}\right)^{1/3}$}                      &\small{Hill radius}                        &&\small{$\mu$}               &\small{$Gm_b$}                                            &\small{}\\                                                      
\small{$\hate$}              &\small{$\v{e}/e$}                                                           &\small{unit eccentricity vector}           &&\small{$\varOmega$}         &\small{$$}                                                &\small{longitude of ascending node}\\                           
\small{$\hath$}              &\small{$\v{h}/h$}                                                           &\small{unit angular momentum vector}       &&\small{$\rm T$}             &\small{$$}                                                &\small{temperature}\\                                           
\small{$\hatq$}              &\small{$\hath \times \hate$}                                                &\small{}                                   &&\small{$\lambda_{\rm mfp}$} &\small{$\frac{\mu_{\rm mol}\ m_H}{\rho\sigma_{\rm coll}}$}&\small{mean free path}\\                                        
\small{$\v{\Omega}_1$}       &\small{$$}                                                                  &\small{spin of primary}                    &&\small{$T_{\rm out}$}       &\small{$2\pi/n_{\rm out}$}                                &\small{period of outer binary}\\                                
\small{$\v{\Omega}_2$}       &\small{$$}                                                                  &\small{spin of secondary}                  &&\small{$\mu_{\rm mol}$}     &\small{$$}                                                &\small{mean molecular weight}\\                                 
\small{$\mu_r$}              &\small{$\frac{m_1m_2}{m_1+m_2}$}                                            &\small{reduced mass}                       &&\small{$m_H$}               &\small{$$}                                                &\small{atomic mass unit}\\                                      
\small{$\mathcal{I}_1$}      &\small{$\frac{2}{5}m_1\left(R_{x1}^2+R_{y1}^2\right)$}                      &\small{inertia moment of primary}          &&\small{$\sigma_{\rm coll}$} &\small{$$}                                                &\small{collisional cross section}\\                                
\small{$\mathcal{I}_2$}      &\small{$\frac{2}{5}m_2\left(R_{x2}^2+R_{y2}^2\right)$}                      &\small{inertia moment of secondary}        &&\small{$\mu_{\rm visc}$}    &\small{$$}                                                &\small{dynamical viscosity}\\                                   
\small{$T$}                  &\small{$2\pi/n$}                                                            &\small{period of inner binary}             &&\small{$c_s$}               &\small{$$}                                                &\small{sound speed}\\                                           
\small{$R_x$}                &\small{$$}                                                                  &\small{principal semiaxis of ellipsoid}    &&\small{$P$}                 &\small{$\rho c_s^2$}                                      &\small{pressure}\\                                                 
\small{$R_y$}                &\small{$$}                                                                  &\small{principal semiaxis of ellipsoid}    &&\small{$\Sigma$}            &\small{$\int \rho dz$}                                    &\small{column density}\\                                             
\small{$R_z$}                &\small{$$}                                                                  &\small{principal semiaxis of ellipsoid}    &&\small{$p$}                 &\small{$$}                                                &\small{power law of column density}\\  
\small{$R$}                  &\small{$(R_xR_yR_z)^{1/3}$}                                                 &\small{equivalent radius of ellipsoid}     &&\small{$\rm Ma$}            &\small{$|\Delta v|/c_s$}                                  &\small{Mach number}\\                  
\small{$J_2$}                &\small{$\frac{1}{10}\frac{R_x^2+R_y^2-2R_z^2}{R^2}$}                        &\small{quadrupole potential}               &&\small{$\rm Kn$}            &\small{$\frac{\lambda_{\rm mfp}}{2R}$}             &\small{Knudsen number}\\                                             
\small{$\tau_1$}             &\small{$\frac{8}{3C_D}\frac{R_1}{|\Delta \v{v}|}\frac{\rho_\bullet}{\rho}$} &\small{drag time of the primary}           &&\small{$\tau_w$}            &\small{$\frac{\tau_1\tau_2}{\tau_1-\tau_2}$}              &\small{wind drag timescale}\\                         
\small{$\tau_2$}             &\small{$\frac{8}{3C_D}\frac{R_2}{|\Delta \v{v}|}\frac{\rho_\bullet}{\rho}$} &\small{drag time of the secondary}         &&\small{$\tau_m$}            &\small{$\tau_{\rm eff}$}                                  &\small{orbital drag timescale}\\       
\small{$\tau_{\rm eff}$}     &\small{$\frac{\tau_1\tau_2(m_1+m_2)}{\tau_2m_2+\tau_1m_1}$}                 &\small{effective drag time of binary}      &&\small{$e_{\rm out}$}       &\small{$|\v{E}|$}                                         &\small{eccentricity of outer binary}\\ 
\small{$\rho_\bullet$}       &\small{$$}                                                                  &\small{internal density}                   &&\small{$\v{x}$}             &\small{$x\hatx+y\haty+z\hatz$}                            &\small{local Hill coordinates}\\       
\small{$G$}                  &\small{$$}                                                                  &\small{gravitational constant}             &&\small{$\v{x}_{\rm cart}$}  &\small{$x_c\hate+y_c\hatq+z_c\hath$}                      &\small{Cartesian coords. on orbital frame}\\   
\small{$n_{\rm out}$}        &\small{$\sqrt{\frac{GM_\sun}{a_{\rm out}^3}}$}                              &\small{mean motion of outer binary}        &&\small{$\v{x}_{\rm cyl}$}   &\small{$\vt{R}_h(-f)\v{x}_{\rm cart}$}                    &\small{Cylindrical coords. on orbital frame}\\ 
\small{$v_{\rm out}$}        &\small{$n_{\rm out} a_{\rm out}$}                                           &\small{circular velocity of outer binary}  &&\small{$\v{E}$}             &\small{$$}                                                &\small{eccentricity vector of outer orbit}\\   
\small{$a_1$}                &\small{$\frac{m_2}{m_1+m_2}a$}                                              &\small{semimajor axis of primary}          &&\small{$Q$}                 &\small{$$}                                                &\small{tidal dissipation factor}\\             
\small{$a_2$}                &\small{$\frac{m_1}{m_1+m_2}a$}                                              &\small{semimajor of secondary}             &&\small{$\v{s}_1$}           &\small{$$}                                                &\small{distance origin-primary}\\              
\small{$v_1$}                &\small{$na_1$}                                                              &\small{circular velocity of primary}       &&\small{$\v{s}_2$}           &\small{$$}                                                &\small{distance origin-secondary}\\            
\small{$v_2$}                &\small{$na_2$}                                                              &\small{circular velocity of secondary}     &&\small{$\v{r}_1$}           &\small{$-\frac{m_2}{m_1+m_2}\v{r}$}                       &\small{distance barycenter-primary}\\          
\small{$\eta$}               &\small{$$}                                                                  &\small{sub-Keplerian parameter}            &&\small{$\v{r}_2$}           &\small{$\frac{m_1}{m_1+m_2}\v{r}$}                        &\small{distance barycenter-secondary}\\        
\small{$u$}                  &\small{$\eta v_{\rm out}$}                                                  &\small{wind velocity}                      &&\small{$s_{\rm cm}$}        &\small{$$}                                                &\small{distance barycenter-origin}\\           
\small{$u_{\rm eff}$}        &\small{$u \frac{(\tau_2-\tau_1)(m_1+m_2)}{(m_2\tau_2+m_1\tau_1)}$}          &\small{effective wind}                     &&\small{$M_{\rm out}$}       &\small{$n_{\rm out} t$}                                   &\small{mean anomaly of outer orbit}\\          
\small{$\rho$}               &\small{$$}                                                                  &\small{gas density}                        &&\small{$t_{\rm kozai}$}     &\small{\eq{eq:kozai-timescale}}                           &\small{timescale of Kozai-Lidov cycle}\\       
\small{$C_D$}                &\small{\eq{eq:cd}}                                                          &\small{drag coefficient}                   &&\small{$f_H$}               &\small{$a/R_H$}                                           &\small{semimajor axis in Hill radii}\\             
\small{$\Rey$}               &\small{$2 R \rho |\Delta \v{v}|/\mu_{\rm visc}$}                     &\small{Reynolds number past the object}    &&\small{$b_c$}               &\small{$$}                                                &\small{impact parameter}\\             
\small{$n$}                  &\small{$\sqrt{\frac{Gm_b}{a^3}}$}                                           &\small{mean motion of inner binary}        &&\small{$f_D$}               &\small{$b_c/\bar{D}$}                                     &\small{impact parameter in binary diameters}\\             
\small{$v_{\rm esc}$}        &\small{$\sqrt{2G(m_1+m_2)/R}$}                                       &\small{escape velocity}                    &&\small{$\nu$}               &\small{$\frac{16}{3}\frac{n}{n_{\rm out}}$}               &\small{normalized ratio of mean motions}\\             
\small{$q$}                  &\small{$a(1-e)$}                                                            &\small{pericenter distance}                &&\small{$\beta$}             &\small{$\frac{16}{3\nu^2}$}                               &\small{normalized ratio of mean motions}\\             
\small{$\mu_{\rm b}$}        &\small{$$}                                                                  &\small{rigidity}                           &&\small{$M_\beta$}           &\small{$\beta M$}                                         &\small{scaled mean anomaly}\\             
\small{$\v{r}$}              &\small{$$}                                                                  &\small{separation of inner binary}         &&\small{$\Psi$}              &\small{$\Omega-\nu M_\beta$}                              &\small{modified longitude of ascending node}\\             
\small{$\bar{D}$}            &\small{$\left(\frac{8}{\pi}\frac{m_1+m_2}{\rho_\bullet}\right)^{1/3}$}      &\small{effective combined diameter}        &&&&\\\hline
\end{tabular}
\end{center}
\end{table*}

\subsection{Nebular Drag}

Gas drag enters the equations as the dissipation parameters
$V_d$ and $W_d$ in eccentricity and angular momentum,
respectively. Notice that while $W_i$ are related to
spin-orbit coupling, with the orbit and spin angular momenta 
equations having equal and opposite terms, $W_d$
has no such symmetry. This is because the angular momentum taken from the orbital
motion by the drag is not conserved by converting it into rotational
angular momentum, but given to the nebular gas. 

To find the values of $V_d$ and $W_d$, we work out the orbital solution in the presence of drag, which is shown
in detail in appendix~\ref{app:dragsolution}. The solution,  despite a lengthy and laborious derivation, turns out to be remarkably
simple. If $a_0$, $e_0$, and $h_0$ are the initial semimajor axis,
  eccentricity and angular momentum of the inner binary, the solution
  at a time $t$ is given by 

\beqn
\tilde{a} &=& a_0\ee^{-2t/\teff} \label{eq:solution_drag_a}\\
\tilde{e} &=& e_0 \label{eq:solution_drag_e}\\
\tilde{h} &=& h_0\ee^{-t/\teff} \label{eq:solution_drag_h}
\eeqn

\noindent where tilde represents
average over the solar orbit, and 

\beq
\teff = \frac{\tau_1\tau_2 (m_1+m_2)}{\tau_2m_2 +\tau_1m_1}
\label{eq:teff-equation}
\eeq

\noindent is the effective drag time. Here $\tau_1$ and $\tau_2$ are
  the drag times on the primary and secondary, respectively. The eccentricity is constant, while the angular momentum decays
exponentially in an e-folding time equal to the effective drag
time. As consequence, the energy decays at twice the rate of the
angular momentum. Equations \eq{eq:solution_drag_e} and
\eq{eq:solution_drag_h} lead to the coefficients

\beqn
V_d &=& 0,\\
W_d &=& -\frac{1}{\teff}.
\eeqn

\subsection{Parameters of MU69}

We consider that MU69 was a binary system that came into contact, the
lobes being the primary and secondary masses $m_1$ and $m_2$. The
dimensions of the lobes denote a volume ratio of
roughly 2:1. The equivalent radius of spheres of same volume are
$R_1 = 7.8$\,km and $R_2 = 6.4$\,km. The parameter $J_2$ for each
lobe, assuming a homogeneous triaxial ellipsoid \citep{Scheeres94}, is given by 

\beq
J_2 = \frac{1}{10}\frac{\left(R_x^2+R_y^2-2R_z^2\right)}{R^2}
\eeq

\noindent where $R_x$, $R_y$, and $R_z$ are the principal
semi-axes. For the lobes of MU69, the observed values of $R_x$ ,$R_y$, and
$R_z$ lead to $J_2$ = 0.26 and 0.14 ($\pm 0.07$) for the larger and smaller lobes,
respectively. Assuming an internal density of $\rho_\bullet$ = 0.5
g/cm$^3$, the masses are $m_1 =
1.01 \times 10^{18}$g and $m_2 = 5.45 \times 10^{17}$ g. At the
  distance of $a_{\rm out}=$45\,AU, the center of mass orbits the Sun at the velocity of
$v_{\rm out} = n_{\rm out} a_{\rm out} \approx $  4.5\,km/s, where
  $n_{\rm out}$ is the mean motion of the heliocentric orbit. The Hill radius of the combined masses is 
$R_H=a_{\rm out} [m_b/(3M_\sun)]^{1/3}$, where $m_b=m_1+m_2$ is the sum of
the masses of the primary and secondary and $M_\sun$ is the solar mass. Substituting the masses
obatined above yields $R_H \approx 2.8\times 10^{-4}$\,AU,
or $4.3\times 10^4$\,km. 

For a representative semimajor axis $a= 0.1 R_H \approx 4300 \ {\rm
  km}$ the period is $T=5.6 \, {\rm  yr}$. The semimajor axes of the
orbits around the barycenter and respective
circular orbital velocities are $a_2 = 2785 \,{\rm km}$, $a_1 = 1505\,{\rm km}$,
and $v_2 = 10 \, {\rm cm\,s^{-1}}$, $v_1 = 5.5 \, {\rm cm\,s^{-1}}$.

These orbital velocities are very small{\footnote{In fact small enough that
  impacts, perturbations by other KBOs, or other dynamical effects
  could easily ionize the binary. This is either an indication that
  these effects did not play a significant role, or that the lobes
  formed at closer separation, or both. We cannot assess the latter,
  but that MU69 is not significantly cratered indeed points to a 
  low-collision environment, as expected for the Kuiper belt.}}. The velocity of the center of
mass around the Sun, $v_{\rm out} = 4.5$ km/s, is over $40\,000$ times
larger. For the minimum-mass Solar nebula
\citep[MMSN, ][]{Weidenschilling77,Hayashi81} at 45\,AU, the gas is
sub-Keplerian by $n_{\rm gas} = n_{\rm out} (1-\eta)$,
where $n_{\rm out}$ is the Keplerian value and  $\eta \sim 0.01$. The wind that the binary is experiencing is then 

\beq
u = \eta v_{\rm out} \approx 50 \, {\rm m\,s^{-1}}.
\eeq 

\noindent i.e., about 500  times faster than the circular velocity of the binary. The sub-Keplerian wind cannot in principle be ignored. We find in appendix~\ref{app:dragsolution} that in the reference frame of
motion around the primary, the effective wind is 

\beq 
\v{u}_{\rm eff} = \v{u} \frac{(\tau_2-\tau_1)(m_1+m_2)}{(m_2\tau_2+m_1\tau_1)}.
\label{eq:equiv-wind}
\eeq

\noindent The model is fully specified if the drag times
are known. These are considered in the next section.

\subsubsection{Drag time}
\label{sect:drag}

Solid particles and gas exchange momentum due to interactions that
happen at the surface of the solid body. The many processes that can
occur are generally described by the collective name of ``drag'' or
``friction''. Consider a solid body of cross section $\sigma$
travelling through a fluid medium of uniform density $\rho$ with
velocity $\v{v}$ with respect to the fluid. In a time interval
$dt$, it sweeps a volume $dV = \sigma v dt$. In the reference
frame of the particle, the gas molecules are travelling 
with velocity $-\v{v}$. If all their momentum is transferred to the particle, the force is

\beq
F_{\rm drag} = \rho dV \frac{d\v{v}}{dt} =-\rho\sigma v \v{v}.
\eeq

In aerodynamics it is usual to define a dimensionless factor $C_D$ that takes into account the deviations from this idealized picture

\beq
F_{\rm drag} = -\frac{1}{2}C_D\rho\sigma v \v{v}.
\label{eq:Fdrag}
\eeq

The factor half comes in because it is common to define the drag force
in terms of kinetic energy instead of momentum. Considering spheres of
radius $R$, their cross section is $\sigma=\pi R^2$; the acceleration
$f_{\rm drag}$ in the equation of motion is found upon dividing $F_{\rm
  drag}$ by the mass of the object $m=4 \pi R^3\rho_\bullet/3$, where $\rho_\bullet$
is the material density

\beq
f_{\rm drag} = -\left(\frac{3\rho C_D  v }{8R\rho_\bullet}\right)\v{v}.
\eeq

The quantity in parentheses has dimension of time$^{-1}$, defining the
drag time of an embedded object

\beq
\tau \equiv \frac{8}{3C_D} \frac{R}{v} \frac{\rho_\bullet}{\rho}.
\label{eq:dragtime-equation}
\eeq

The drag time represents the timescale within which the
object couples to the gas flow. The parameter $C_D$ can be calculated
from first principles or derived from experiments, depending on the
drag regime of interest. When the object radius exceeds the mean free 
path of the particle, the approximation of ballistic collisions ceases
to apply and the frequent intermolecular collisions lead to the
emergence of viscous behaviour. It is a well-known result that ideal 
fluids exert no drag (d’Alembert’s paradox). When the kinematic
viscosity $\mu_{\rm visc}$ is considered, 
Stokes drag law on a large sphere (large meaning bigger
than the mean free path of the gas) is recovered

\beq
F_{\rm drag}^{\rm (Stk)} = -6\pi \mu_{\rm visc} R \v{v},
\label{eq:fstokes}
\eeq

\noindent a lengthy proof of which can be found in Landau \&
  Lifshitz (1987). Dividing \eq{eq:fstokes} by the mass of the object and
  expressing it in the form of \eq{eq:Fdrag}, one finds 

\beq
C_D^{\rm (Stk)}  = \frac{12\mu_{\rm visc}}{R\rho v} = \frac{24}{\Rey}
\label{eq:cd_stk}
\eeq

On obtaining this equation, the inertia of the fluid is neglected, so
it only holds for low Reynolds numbers. Empirical corrections to
Stokes' law were worked out (e.g., Arnold 1911, Millikan 1911,
Millikan 1923), but a general case derived from first principles is
difficult to obtain. The major complication resides at the boundary
layer immediately over the surface of the particle, where the velocity
of the viscous fluid has to be zero. If the fluid has inertia, a sharp
velocity gradient develops in the flow past the object as the velocity
goes to zero at the solid surface. At this boundary layer, the viscous
term is important even at high Reynolds numbers 
(Prandtl 1905). It can be seen experimentally that in such cases, 
the flow past the particle develops into a turbulent wake (von Karman
1905), with drag coefficients much larger than those predicted by Stokes law.

Experiments with hard spheres \citep{Cheng09} show 
that the drag coefficient $C_D$ for large objects and valid in the
range $\Rey < 2\times 10^5$ can be fit by
the following empirical formula \citep{PeretsMurrayClay11}

\beq
C_D = \frac{24}{\Rey}\left(1+0.27\right)^{0.43} +0.47\left[1-\exp{(-0.04\Rey^{0.38})}\right].
\label{eq:cd}
\eeq

\noindent 
The value of $C_D$ varies
non-monotonically, being $C_D\approx26$ for $\Rey=1$, reaching a
minimum of $C_D\approx 0.25$ at $\Rey=10^3$ and rising to $C_D\approx 0.45$ at $\Rey=10^5$. The drag times for the pre-merger lobes of MU69 are $\tau_1=2.87\times 10^7$\,yr and $\tau_2 = 2.00 \times
10^7$ yr (details of the calculation are shown in appendix
~\ref{app:dragtime}). The system can thus be modeled as going through
an effective headwind $u_{\rm eff} = 166 \, u \approx 25$\,m/s with effective friction
time $\tau_{\rm eff} = 2.24\times 10^7$\,yr. 

We highlight that because $F_{\rm drag} \propto C_D \Rey^2$, in
  the low Reynolds number regime, with $C_D$ given by \eq{eq:cd_stk},
  the drag force is linear with velocity ($\Rey \propto v$). Conversely, in the regime of high Reynolds number ($\Rey
\approx 10^3)$, $C_D$ asymptotes to a constant value, and
\eq{eq:Fdrag} becomes quadratic with velocity. This is the regime of
turbulent, or ram pressure, drag. The different regimes are shown in
the left panel of \fig{fig:drag}, where the y-axis is $C_D \Rey^2$. As seen in the figure, linear drag is valid up to $\Rey
\approx 1$, and quadratic beyond  $\Rey \approx 1000$ with a smooth
transition in between. For MU69 at 45\,AU in the MMSN, the
Reynolds number is $\Rey \approx 10$, placing it much closer to linear
(viscous) than to quadratic (turbulent) drag. This distinction has
profound consequences for the orbital evolution of the pre-merger
lobes of MU69. 

\section{Results}
\label{sect:results}

\subsection{Inability of drag alone to lead to contact in the Kuiper belt}

The drag times for bodies of the size of MU69 as a function of
distance are shown in the upper panels of \fig{fig:drag_linearvsquadratic}. In these curves the
Reynolds number is calculated by having the velocity being the sum of
the wind and the orbital velocity (at a representative distance of 0.1
$R_H$ from the central object with mass equal to the combined mass of
MU69). The left panel is the MMSN, and the right panel a solar
  nebula ten times more massive than the MMSN. The different lines are
different values for the wind velocity. The red solid line corresponds
to standard pressure gradients, yielding wind velocities of $\approx$
50\,m/s. The blue solid line is no wind, corresponding to a pressure
maximum. The other lines represent reduced or enhanced pressure
gradients, for comparison.

In the outer disk the Reynolds number is low enough that
viscous linear drag ensues and the presence of the wind does
not matter. This is because if the drag is linear, the angular momentum gained in half an orbit is exactly
  lost in the other half whereas for
quadratic drag the influence of the wind does not cancel out exactly
when averaged over an orbit \citep[cf. Sect 3.2 of ][]{PeretsMurrayClay11}; as a result, in the inner disk where the
drag is quadratic, the influence of the wind in the drag time is
dominant. The upper panels of \fig{fig:drag_linearvsquadratic} shows that the wind-enhanced orbital drag alone, with no Kozai cycles, is
able to collapse a MU69-like binary in the asteroid belt in the timeframe of the
lifetime of the disk, but not in the Kuiper belt, either in the
  MMSN or in a nebular ten times more massive.

The lower panels of \fig{fig:drag_linearvsquadratic}
  show drag time (color coded) and Reynolds numbers (solid lines) as a function of distance in AU and
  body radius in km, for a wind speed of 50\,m/s. The regimes of linear (Re $<$ 1) and
  quadratic drag (Re $>$ 10$^3$) are shown as thicker solid lines and
  arrows. The pre-merger lobes of MU69 are shown as red dots in both
  plots: as seen in the figures, they lie in the transition between
  linear and quadratic drag. In the MMSN the Reynolds number is about
  10, closer to linear, with drag times of $\approx$20 Myr; for the more massive model (10x MMSN) the
  Reynolds number is closer to 100. In this more massive case the drag
  time is $\approx$10 Myr; the effect of the wind, even though closer to quadratic than to linear,
  is still not enough to lower the drag time to values within the
  lifetime of the nebula.

For the Kuiper belt, the timescales of the problem are, in decreasing order:
dynamical ($n$, where $n$ is the orbital mean motion);
Coriolis force ($n_{\rm out} \approx 10^{-2}n$); centrifugal
($n_{\rm out} ^2\approx 10^{-4}n$); wind $(u/\tau \approx 10^{-5}n)$;
orbital drag ($1/\tau \approx 10^{-7}n$); shear ($n_{\rm out} /\tau
\approx 10^{-9}n$). We plot in \fig{fig:incl00} the $N$-body evolution of a binary of
point masses, subject to gas drag, with initial inclination
$I$=0$\degree$, 60$\degree$, and 90$\degree$, for a range of
eccentricities. The lines shown are box-averages over a solar period. 

The zero inclination curves show the predicted behavior
of semimajor axis, eccentricity and angular momentum, averaged over a
solar orbit. The system has an exponential decay of angular
momentum and semimajor axis, with
  e-folding time defined by the drag time $\tau$. This rate of decay
is enough to lead to collapse in the asteroid belt, where the
quadratic drag aided by the wind brings $\tau$ to 0.1\,Myr
timescales. Yet, at the Kuiper belt, as discussed above, the timescales remain of the order of 10\,Myr,
hindering collapse. 

\begin{figure*}
  \begin{center}
    \resizebox{\textwidth}{!}{\includegraphics{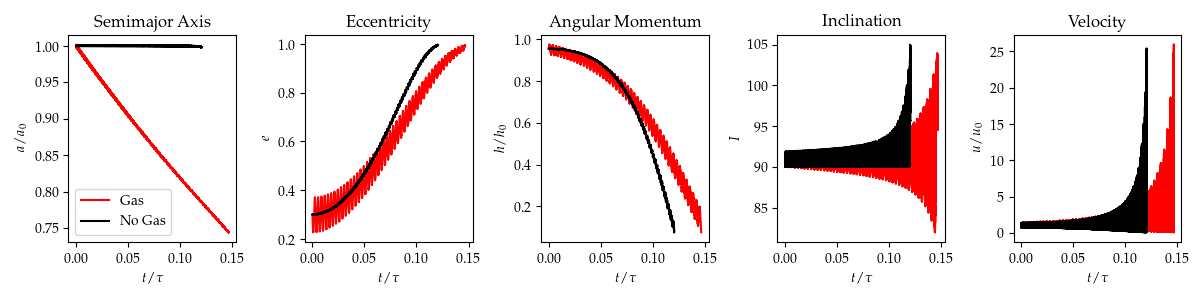}}
\end{center}
\caption{Comparison between the model shown for $90\degree$ on \fig{fig:incl00} and
  the same model but switching off nebular drag. The evolution is mainly
  driven by the Kozai oscillations, with gas playing a secondary role and the wind
  driving the eccentricity and angular momentum oscillation. Notice
  how nebular drag makes the inclination flip between prograde and
  retrograde, and how the velocity increases to 20-25 times the initial
  orbital velocity as the eccentricity approaches unity.} 
\label{fig:kozai}
\end{figure*}

\subsection{Kozai-driven collapse in the Kuiper belt}

The $I=60\degree$
plots show oscillations of eccentricity and inclination; we measure
that these oscillations conserve the Kozai constant $H_k$, characterizing Kozai-Lidov oscillations. Still the
excursions into high eccentricity are not enough to make the orbit
grazing (separation less than 30\,km) even for initial eccentricity
$e_0=0.9$ over 10\,Myr.

The situation changes for $I=90\degree$. Now, for moderate initial 
eccentricities, the eccentricity shoots to unity in very short
timescales, and the orbit becomes grazing,
reaching separation $r=30$\,km in timescales of the order of
$0.1\tau$. The collapse happens essentially at constant energy. The orbit is
forced from initially circular into a progressively elongated ellipse
and finally into a straight line, leading to contact (\fig{fig:trajectory}). 

Since the orbit is Keplerian, we can analytically calculate the
velocity at contact. For a head-on collision at pericenter 

\beq
v = \sqrt{\frac{G(m_1+m_2)}{a} \left(\frac{1+e}{1-e}\right)}
\eeq

\noindent where $G$ is the gravitational constant. Writing $v_{\rm esc} = \sqrt{2G(m_1+m_2)/R}$ for the escape velocity ($R$
being the effective radius of MU69), we can write this equation as 

\beq
\frac{v}{v_{\rm esc}} = \sqrt{\frac{R}{q} \left(1-\frac{q}{2a}\right)}
\eeq

\noindent where $q$ is the pericenter distance. If contact happens along the principal
axes, $q = (R_{x1} + R_{x2})/2 \ll a$ (the subscripts 1 and 2 refer to primary and secondary bodies), and we have

\beq
\frac{v}{v_{\rm esc}} = \frac{(R_{x1}R_{y1}R_{z1}+R_{x2}R_{y2}R_{z2})^{1/6}}{(R_{x1}+R_{x2})^{1/2}}.
\eeq

For the parameters of MU69, this yields $v \approx 0.71 v_{\rm esc}$, i.e.
about 70\% reduction compared to the escape velocity. The escape velocity depends on the bulk density of MU69, which is not
well constrained. The velocity at contact is 

\beq
v \approx 4.2 \, {\rm m\,s^{-1}} \ \sqrt{\frac{\rho_{\bullet}}{0.5\, {\rm g\,cm^{-3}} } }.
\eeq

An internal density of 0.3\,g\,cm$^{-3}$ leads to contact at velocity
3.3\,m\,s$^{-1}$. Collision velocities in the range $2-3$ m\,s$^{-1}$
happen for internal densities
in the range $0.12 - 0.25$ g\,cm$^{-3}$. \fig{fig:kozai} shows a comparison of the
same simulation with nebular drag switched on and off. Gas
drag plays a secondary role: the evolution is driven mostly by
Kozai-Lidov cycles. Just how secondary the role of nebular drag is is
explored in the next section with our Kozai-Tides-J2-drag model. 

\subsection{Gas-enhanced Kozai}

The orbital integrations shown in \fig{fig:incl00} treat the bodies
as point masses, which are gravitational monopoles and impervious to
tides. To relax this approximation, we make use of the
orbit-integrated KTJD model described in \sect{sect:model}. For the tidal model we use rigidity $\mu_{\rm b} =
\xtimes{4}{10}$ g\,cm$^{-1}$\,s$^{-2}$ and
tidal dissipation quality $Q=100$, as typically
assumed for icy bodies \citep{RagozzineBrown09}. 

Considering small initial eccentricity, the maximum eccentricity induced by Kozai is \citep{PeretsNaoz09}

\beq
e_{\rm max} = \sqrt{1 - \frac{5}{3} \cos^2 I_0}.
\eeq

\noindent We want to find the range of inclinations for which the orbit is
grazing, i.e., $r < R_b$, where $R_b=R_1+R_2$ is the sum of the radii
of the object. The separation being $r=q\equiv a(1-e)$, the critical
inclination is, considering small initial eccentricities, 

\beq
I_{\rm crit} = \cos^{-1}\left(\sqrt{\frac{6}{5}\frac{R_b}{a}}\right).
\label{eq:icrit}
\eeq

The timescale of the Kozai oscillation is given by \cite{Kiseleva+98}

\beq
t_{\rm kozai} = \frac{2 T_{\rm out} ^2}{3\pi T} \frac{\left(m_1+m_2+M_\sun\right)}{M_\sun}\left(1-e_{\rm out} ^2\right)^{3/2}
\label{eq:kozai-timescale}
\eeq

\noindent where $T_{\rm out} $ and $e_{\rm out} $ are the period and the
eccentricity of the orbit around the Sun, respectively. For $T_{\rm out} $=350\,yr, $T$=5\,yr, and $e_{\rm out} =0.04$, yields a timescale for contact via Kozai of 
$t_{\rm kozai}\approx 5000$\,yr.

\begin{figure*}
  \begin{center}
    \resizebox{\textwidth}{!}{\includegraphics{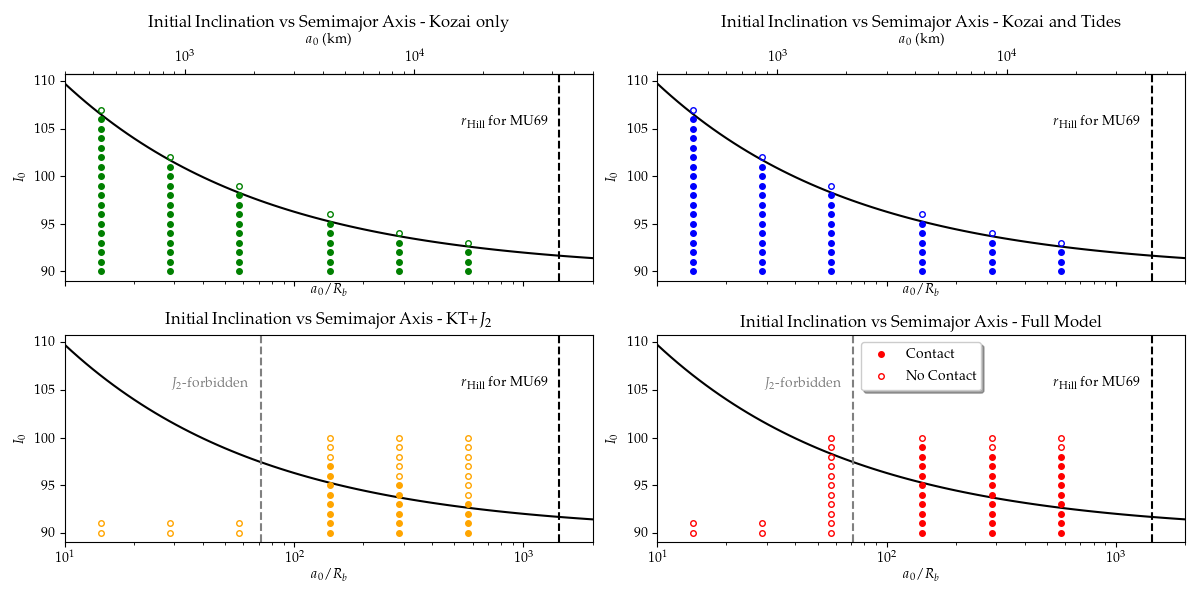}}
\end{center}
\caption{Mapping of the parameter space of semimajor axis and
  inclination (for fixed eccentricity $e_0=0.1$)  over which contact
  happens. The upper left plots refers to a model with only Kozai
  oscillations and no dissipation, $J_2$ or drag. For this model, the
  predicted critical inclination is shown as the solid line. The dots
  represent the simulations we ran. Filled dots represent contact,
  open dots no contact. The model finds excellent agreement with the
  prediction (only $I_0>90\deg$ is shown but the results
  are symmetric for prograde orbiters). The upper right plot shows the model of Kozai and
  dissipation. Because spin-orbit coupling is weak, not much
  distinction is seen between this and the model of Kozai only. The
  lower left plot shows a model with Kozai, tides, and $J_2$. The
  quadrupole disrupts Kozai oscillations inside of 0.05 $R_H$, but it
  extends slightly the critical inclination. The lower right plot
  shows the full model with Kozai, tides, $J_2$ and drag. The $J_2$-forbidden region still 
  exists, but outside this, the range of inclination over which contact
  happens is significantly increased.}  
\label{fig:critical_inclination}
\end{figure*}

We ran models with initial eccentricity $e_0=0.1$ and explored the parameter
space of semimajor axis and inclination. \fig{fig:critical_inclination} shows inclination vs semimajor axis plots
with the results of the integrations. The lower $x$-axis is the semimajor axis
in units of $R_b$, which we take to be 30\,km, approximately the principal
axis of MU69. The upper $x$-axis shows the semimajor axis in km. A black
dashed line shows the Hill radius of MU69. We use an upper limit of
$a=0.4 R_H$ since beyond this semimajor axis the orbits are heavily disrupted
by the solar tide. Also, if the eccentricity goes near unity for $a
\geq 0.5 R_H$, the apocenter is outside the Hill sphere and
the binary is ionized. The semimajor axes sampled are
$a/R_H = 0.01, 0.02, 0.04, 0.1, 0.2$, and $0.4$.

The different panels show models that consider only Kozai (K, upper-left panel, in green); Kozai and
tides (KT, upper-right panel, in blue); Kozai, tides, and $J_2$ (KTJ,
lower left panel, in orange), and finally the full model of Kozai,
tides, $J_2$, and drag (lower right panel, in red). Each integration
was done until 10\,Myr. Simulations where contact happened are shown as filled circles, whereas empty circles
denote no contact. We consider only retrograde inclinations. 

The solid line in the plots shows the critical inclination for contact
given by \eq{eq:icrit}. As seen in the upper
left plot, the behavior is very well reproduced by the K model. Under the critical
inclination it takes half a Kozai period to achieve contact, as the maximum
eccentricity brings the pericenter inside the primary. The
Kozai oscillations are periodic and regular, so above the critical
inclination no contact is possible. 

The critical inclination line is also well reproduced by the KT model (upper right), which evidences
how weak spin-orbit coupling is. Indeed, we find that
during the evolution, the spin angular momentum increases by less than
0.1\%. This justifies, a posteriori, our choice of initializing the
spin periods at 15 hrs, the measured rotational period of MU69. 

The situation changes when the permanent quadrupole is
included (lower left). As found by \cite{PorterGrundy12},  
Kozai cycles are thwarted by too strong a $J_2$. Because $J_2$ induces precession, 
it removes the binary from the locked $\dot{\omega}=0$ Kozai
resonance. The behavior is
reproduced in our model, showing a $J_2$-forbidden zone inside the
grey-dashed line at $a/R_H =  0.05$. A slight increase in the occurrence of contact is seen in
the region from $a=0.1$ to $0.4 R_H$. 

\begin{figure*}
  \begin{center}
    \resizebox{\linewidth}{!}{\includegraphics{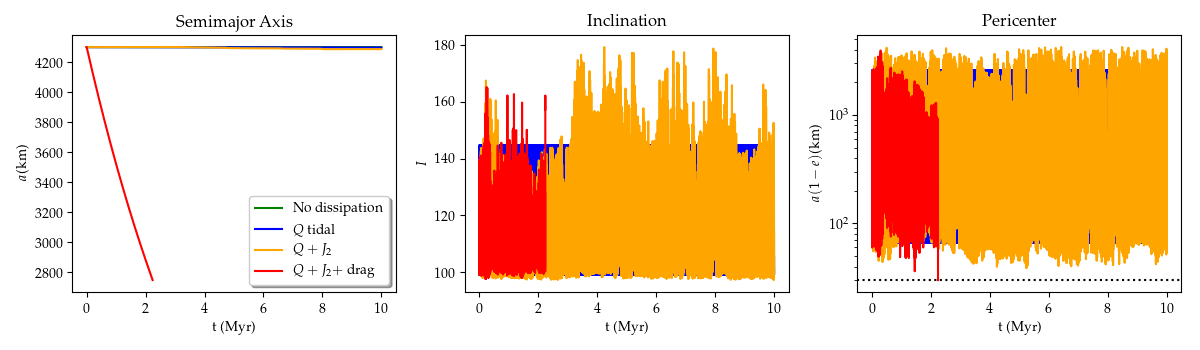}}
\end{center}
\caption{Evolution for $I_0=99\degree$ up to 10\,Myr, for simulations with different
  dynamical terms. The simulation with only the solar terms (green line) leads to
  regular Kozai oscillations at constant semimajor axis. Including the
  induced quadrupole (blue line) has little effect. Inclusion of the permanent quadrupole (orange line) leads to
  irregular Kozai cycles with erratic excursions in inclination and
  eccentricity. Finally, including the orbital drag leads to a fast decay
  of semimajor axis, and eventual contact shortly after 2\,Myr.} 
\label{fig:drag_aided}
\end{figure*}

The full KTJD model (lower right) shows that the inclusion of nebular drag
does not shorten the $J_2$-forbidden zone. The main difference between
KTJ and KTJD is that the occurrence of contact in the $J_2$-allowed
region increases significantly above the critical inclination. \fig{fig:drag_aided} shows the cycles for $I=99\degree$ and $a=0.1
R_H$. The left panel shows semimajor axis, the middle one
inclination, and the right one the pericenter distance. While the simulations K and KT lead to regular cycles
in a well-defined range of eccentricity and inclination at constant semimajor axis, including $J_2$ makes the excursions in
eccentricity and inclination stochastic. However, over 10 million years the
pericenter did not reach 30\,km (dashed line) for the KTJ model. The
inclusion of nebular drag does not seem to have a significant effect in
inclination, but by lowering the semimajor axis over Myr timescales,
it lowers the pericenter distance accordingly when compared to the
model without nebular drag. Assisted by this effect, random variations are
able to bring the binary into contact more easily. 

\begin{figure*}
  \begin{center}
    \resizebox{\textwidth}{!}{\includegraphics{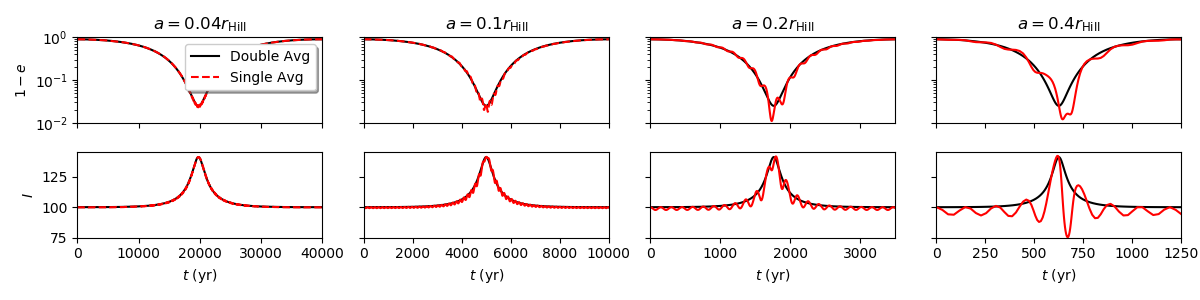}}
\end{center}
\caption{Comparison between the single and double-averaged
  secular approximations, for different values of the semimajor axis,
  shown as a fraction of the Hill radius (0.04, 0.1, 0.2, and
  0.4). The upper plots show the eccentricity, the lower plots show
  the inclination. where the former is averaged over the mean
  anomaly of the inner binary only, and the latter over the mean
  anomalies of the inner and outer binaries. The double average is
  applicable up to 0.1 Hill radii. Beyond that, oscillations 
  on top of the double-averaged solution are seen, increasing
  amplitude with increasing semimajor axis. These oscillations should
  make contact more likely as they bring the eccentricity beyond the
  maximum predicted by the double-averaged model.} 
\label{fig:SinglevsDouble}
\end{figure*}

We caution that our model uses the double-averaged secular approximation, where the motion is averaged in 
mean anomaly of both the inner and the outer binary. We work out in 
Appendix~\ref{app:single-double} that the double-averaged model is
applicable up to $a\approx 0.1 R_H$. We compare in that appendix the prediction of a pure Kozai (no
tides or dissipation) double-averaged model with those of a pure Kozai
single-averaged model, where the motion is averaged over the mean
anomaly of the inner boundary only, resolving the motion of the outer
boundary. As also worked out in the appendix, this approximation is
applicable up to $a\approx 0.3 R_H$, beyond which N-body is
necessary. A comparison between the single-average and double-average
solutions is shown in \fig{fig:SinglevsDouble}, for $a/R_H$=0.04, 0.1,
0.2, and 0.4. Upper plots show the eccentricity, lower plots the
inclination. One Kozai-Lidov cycle is shown for each case. The simulations show that the single-averaged
model has oscillations on top of the double-averaged solution, with
amplitude increasing with semimajor axis. These
extra oscillations bring the eccentricity beyond the maximum
nominal eccentricity predicted by the double-averaged model, and thus
should make contact even more likely. 

A final comment is warranted on the observed inclination of
  MU69, which is $I=99\degree$. For pure quadrupole double-averaged Kozai the
  inclination at contact should be

\beq
\cos I = \cos I_0 \sqrt{\frac{a^2}{R_b}\left(\frac{1-e_0^2}{2a-R_b}\right)}\nonumber
\eeq

\noindent and only a narrow range of the parameter space of
  initial semimajor axis, inclination, and eccentricity would lead to contact at $I=99\degree$. However, because the inclusion of
$J_2$ and drag turns the eccentricity and inclination excursions
stochastic, the equation above is rendered invalid as a predictor of final
inclination. Yet, as seen in the middle panel of \fig{fig:drag_aided}
the general trend still is of $\cos I > \cos I_0$, so the final
inclination should be closer to 180$\degree$ (indeed contact happens in
this model at 160$\degree$). Nevertheless, as shown in the lower panels of
\fig{fig:SinglevsDouble}, the single-averaged model allows for $\cos I <
\cos I_0$ at high eccentricities, including retrograde-prograde flipping, which is not allowed in the double-averaged
approximation. Thus, our model does not allow for more detailed
conclusions on final inclination, but there is indication that the observed $I=99\degree$ inclination of MU69
should be a more likely outcome than the double-averaged model permits. 

\section{Conclusion}
\label{sect:conclusion}

In this work we present a solution for the two-body problem and for
the hierarchical three-body problem with nebular drag, implementing the latter
into a Kozai cycles plus tidal friction model. We divide the nebular
drag into orbital drag and the sub-Keplerian wind that is effected by
the large-scale pressure gradient of the Solar Nebula. The wind is of
the order of 50\,m/s, whereas the orbital velocity of 10\,km bodies is
of the order of 10\,cm/s. The typical drag timescale for 10\,km bodies
is of the order of $10$\,Myr, but for quadratic drag the wind brings
the effective drag timescales down to 0.1\,Myr. For linear drag the
effect of the wind cancels out and the timescale remains 10\,Myr. The
regime of quadratic drag corresponds to distances in the asteroid
belt, whereas regime of linear drag corresponds to distances in the
Kuiper belt. Our model therefore predicts that the asteroid belt
should be significantly depleted of pristine binary planetesimals, as nebular drag is effective in
bringing them to contact. Observations show that the binary fraction
among asteroids is about 15\% \citep{Margot+15} whereas in
the Kuiper belt it can be as high as 40\%
\citep{Noll+08a,Noll+08b,Nesvorny11,Fraser+17}. Unfortunately we cannot draw conclusions from these numbers because the asteroid belt
is highly collisionally evolved and thus the binary population there
is not primordial: only the cold classical population of the Kuiper belt can be used
as a diagnostic for initial binary fraction \citep{MorbidelliNesvorny19}. 

For the Kuiper belt, where the drag timescales
are of the order of 10\,Myr, we find
that Kozai-Lidov oscillations are paramount to achieve contact. If the
inclination is near 90$\degree$, the eccentricity of the inner binary
increases to near unity and the orbits become grazing. The evolution is characterized by decreasing angular momentum at constant
energy, which geometrically means decreasing the semiminor axis while
keeping the semimajor axis constant, eventually collapsing the orbit
into a straight line (\fig{fig:trajectory}). 

The speed at contact is the orbital velocity; if contact happens at pericenter at high
eccentricity, it deviates from the escape velocity only because of the
oblateness, independently of the semimajor axis. For MU69, the oblateness leads to a 30\% decrease in
contact velocity with respect to the escape velocity, the latter scaling with the
square root of the density. For mean densities in the range
0.3-0.5 g\,cm$^{-3}$, the contact velocity should be $3.3-4.2$
m\,s$^{-1}$. 

The timescale for Kozai cycles for MU69 in the range of
semimajor axes of 0.05 to 0.5 Hill radii are at maximum 20\,000 years (and as low as 500 years),
so contact in this model should have happened right after
formation. Considering that the permanent quadrupole $J_2$
prevents Kozai oscillations for semimajor axes shorter than 0.05 Hill
radii, excluding this ``$J_2$ forbidden zone''  confines contact to a narrow window of the
parameter space, in a range of initial inclinations between 85$\degree$
and 95$\degree$. Formation by streaming instability
\citep{Nesvorny+19} results in a broad inclination distribution at
birth; so the narrow 10$\degree$ window means that this model predicts that the fraction of contact binaries should be
about 5\%, which is too on the low end of the observed inclination distribution of KBO binaries \citep{Grundy+11}. 

We find that gas drag significantly alters this picture. The
permanent quadrupole also has the effect of making the Kozai oscillations
stochastic in the range of semimajor axes where they are allowed. This
leads to the possibility that the Kozai cycles, previously regular and
periodic, can now achieve contact by stochastic fluctuations that
nudge the body beyond the allowed region for pure Kozai. Indeed we see
this behavior, but very limited when only the quadrupole
but no gas is included. The stochastic fluctuations push the window of
contact by 1 or 2 degrees beyond the critical inclination of pure Kozai, but no
further. When gas drag is included, the window is pushed to the range
from 80$\degree$ to 100$\degree$. This happens because of a combination of the stochastic
fluctuations caused by $J_2$ and the fact that gas drag is shrinking
the semimajor axis. After a few million years (still within the
lifetime of the disk), the semimajor axis has shrunk enough to bring the contact pericenter within reach of the
stochastic fluctuations. Together, gas drag and $J_2$ can achieve what neither
could in isolation. The synergy widens the window of contact
to over 10\% of the range of inclinations. If the disk is long-lived enough, the window
could be pushed to even higher inclinations. 

We underscore that our solution naturally provides an explanation to one of the main questions posed
by MU69's nature as a contact binary, in constrast to many cold classicals in the 
100\,km range that are detached binaries. If the drag time for MU69 is
of the order of 10 Myr, an object 10 times bigger would have drag
time of 1 Gyr: the effect of nebular drag would be negligible. The
situation for 100\,km bodies is that of the lower left plot of
\fig{fig:critical_inclination}, that depicts Kozai cycles, tides, and
the permanent qudrupole, excluding nebular drag, with a narrower
window of contact.

As limitations of the work, our KTJD model is accurate only up to the
quadrupole approximation. Including the octupole would make the cycles 
even more chaotic \citep{Naoz16}. Also, we use the double-averaged
secular approximation, where the motion is averaged in 
mean anomaly of both the inner and the outer binary. This
approximation underestimates the maximum eccentricity and inclination
range when compared to the single-average approximation (averaging only in the mean anomaly of
the inner binary but resolving the motion of the outer binary) and of
course also compared to the exact solution. Both situations
potentially increase the region of the parameter space over which contact happens, so our
solution may be seen as a conservative lower bound.

\acknowledgements 

We acknowledge discussions with Orkan
  Umurhan, Casey Lisse,  John Spencer, David Nesvorn\'y, and Rebecca Martin. WL  acknowledges support from the NASA Exoplanet Research Program
through grant 16-XRP16\_2-0065. A.N.Y acknowledges support from the
NSF-AAG program through grant 1616929, and from the NASA TCAN
program. A.J. was supported for this work by a Wallenberg Academy Fellow grant from the Knut and Alice Wallenberg
Foundation (grant number 2017.0287). While this paper was being refereed, a work by McKinnon et
  al. was published, that independently came to the conclusion of the
  importance of nebular drag for the orbital evolution of contact
  binaries. However, that work considers only the quadratic drag law,
  which we show here is not the correct regime for the Kuiper belt. 

\bibliographystyle{apj}

\appendix

\section{A: Orbital Solution with Nebular Drag}
\label{app:dragsolution}

\begin{figure}
  \begin{center}
    \resizebox{.6\textwidth}{!}{\includegraphics{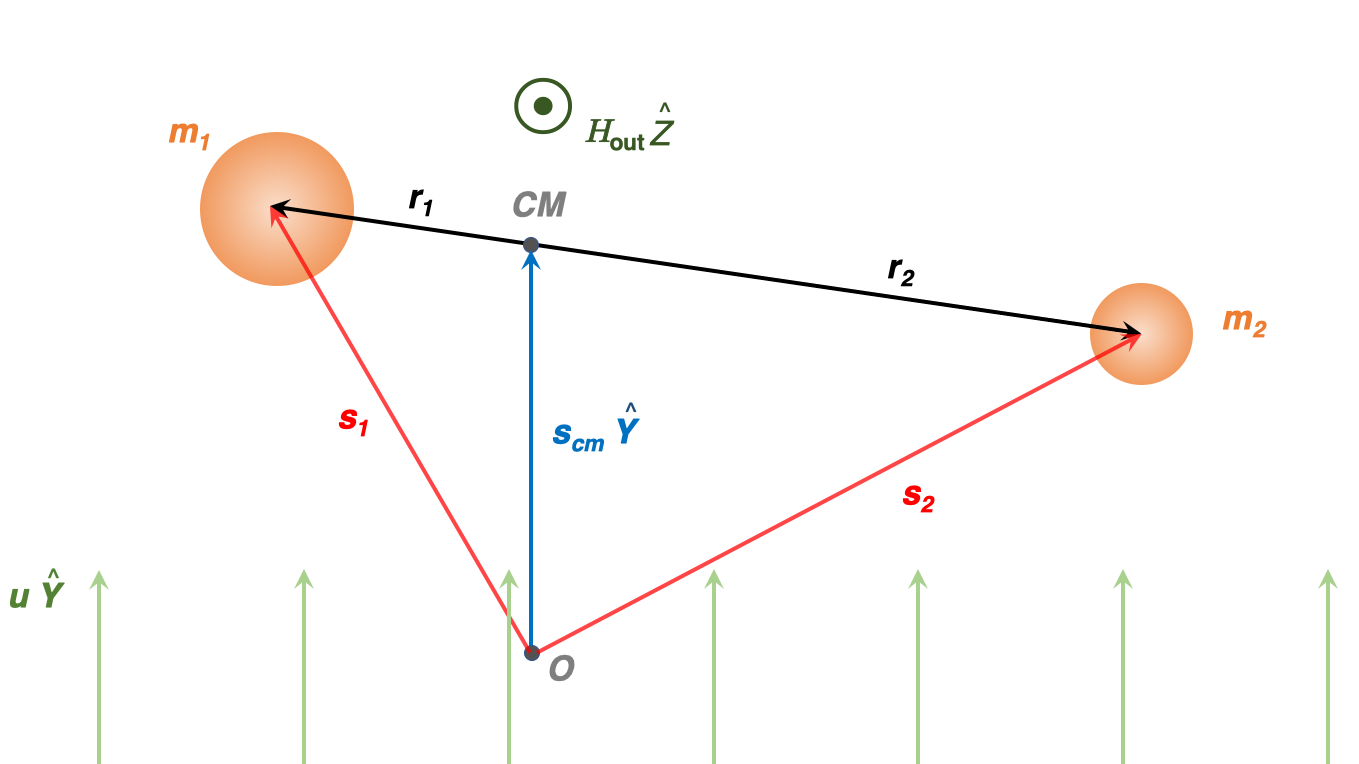}}
\end{center}
\caption{Geometry of the orbital problem with nebular drag. The masses orbit a common center of
  mass at distances $\v{r_1}$ and $\v{r_2}$ from it. We choose an arbitrary origin distant
  $\v{s}$ from the center of mass, and $\v{s_1}$ and $\v{s_2}$ from
  the masses. We align the local Hill coordinate $\haty$ along the
  direction of the wind $\v{u}$, and $\hatz$ along the direction of
  the angular momentum of the orbit around the Sun $\v{H}$.}
\label{fig:geometry}
\end{figure}

We consider a binary system of masses $m_1$ and $m_2$, at distances
$\v{s}_1$ and $\v{s}_2$, respectively, from an arbitrary origin (see \fig{fig:geometry}). The bodies
are immersed in uniform gas and suffer drag as they orbit. The drag
force acts on timescales $\tau_1$ and $\tau_2$, respectively. We 
consider that while the binary's center of mass orbits the Sun at the
Keplerian rate $n_{\rm out} $, the gas moves at sub-Keplerian velocity, leading to a
uniform headwind on the binary with velocity $\v{u}$. We distinguish
between the orthogonal coordinate systems $\hate\hatq\hath$ defined by
the orbit and $\hatx\haty\hatz$, the local Cartesian Hill coordinates
where $\hatx$ points away from the Sun, and $\hatz$ to the angular
momentum vector of the orbit around the Sun. The equations of motion are 

\beqn
\ddot{\v{s}}_1 &=& -2n_{\rm out} \left(\hatz \times
  \dot{\v{s}}_1\right)-Gm_2\frac{\left(\v{s}_1-\v{s}_2\right)}{|\v{s}_1
  -\v{s}_2|^3} +3n_{\rm out} ^2x_1\hatx-\frac{\left(\dot{\v{s}}_1+3/2n_{\rm out}  x_1\haty-\v{u}\right)}{\tau_1}\label{eq:r1}\\
\ddot{\v{s}}_2 &=& -2n_{\rm out} \left(\hatz \times
  \dot{\v{s}}_2\right)-Gm_1\frac{\left(\v{s}_2-\v{s}_1\right)}{|\v{s}_1
  -\v{s}_2|^3}+3n_{\rm out}
^2x_2\hatx-\frac{\left(\dot{\v{s}}_2+3/2n_{\rm out}
    x_2\haty-\v{u}\right)}{\tau_2}\label{eq:r2}
\eeqn

\noindent where $G$ is the gravitational constant. This system can be
reduced to a single particle equivalent as detailed below.  

\subsection{A.1. Single particle equivalent system}

We subtract \eq{eq:r1} from
\eq{eq:r2}, i.e., centering at the primary, and substitute $\v{r}
= \v{s}_2-\v{s}_1$ for the distance between 
the masses. With these operations, the system is 

\beq
\ddot{\v{r}}= -2n_{\rm out} \left(\hatz\times \dot{\v{r}}\right)-\frac{\mu\v{r}}{r^3} +3n_{\rm out} ^2x\hatx- \frac{\dot{\v{s}}_2}{\tau_2}+\frac{\dot{\v{s}}_1}{\tau_1}- \frac{3 n_{\rm out} \haty}{2}\left(\frac{ x_2}{\tau_2}-\frac{x_1}{\tau_1}\right)+\v{u}\left(\frac{1}{\tau_2}-\frac{1}{\tau_1}\right)
\eeq

\noindent where $\mu=G(m_1+m_2)$. The bodies' positions $\v{s}_1$ and $\v{s}_2$ with respect to the origin relate to the barycenter position
$\v{s}_{\rm cm}$ (also with respect to the origin) and the bodies' positions
relative to the barycenter $\v{r}_1$ and $\v{r}_2$ by 

\beq
\v{s}_1  = \v{r}_1 + \v{s}_{\rm cm} \quad {\rm and} \quad \v{s}_2 =\v{r}_2 + \v{s}_{\rm cm} \label{eq:cm}
\eeq

\noindent which, given the definition of the barycenter, yields 

\beq
\v{r}_1 = -\frac{m_2}{m_1 + m_2} \v{r}; \quad \v{r}_2 = \frac{m_1}{m_1 + m_2}  \v{r},\label{eq:radii}
\eeq

\noindent now substituting $\v{r}_1$ and $\v{r}_2$ as given by
\eq{eq:cm} we have 

\beq
\ddot{\v{r}}= -2n_{\rm out} \left(\hatz
  \times\dot{\v{r}}\right)-\frac{\mu\v{r}}{r^3}+3n_{\rm out} ^2x\hatx
-\frac{\v{\dot{r}}}{\tau_m} -\frac{3/2n_{\rm out}
  x\haty}{\tau_m}+\frac{\v{u}-\dot{\v{s}}_{\rm cm}}{\tau_w} 
\label{eq:v21}
\eeq

\noindent with the drag timescales

\beq
\tau_m = \frac{\tau_1\tau_2 \ (m_1+m_2)}{\tau_2m_2 + \tau_1m_1} 
\label{eq:taueff}
\eeq

\noindent and 

\beq
\tau_w = \frac{\tau_1\tau_2}{\tau_1-\tau_2}.
\eeq

\noindent Notice that \eq{eq:v21} is not yet a single particle
equation, as it depends on the motion of the center of mass of the
binary. We consider $u\gg\dot{s}_{\rm cm}$ to drop the last term. With this
approximation, we can further simplify \eq{eq:v21} by writing 

\beq
\frac{\v{\dot{r}}}{\tau_m} - \frac{\v{u}}{\tau_w} =
\frac{\v{\dot{r}} - \v{u}\tau_m\tau_w^{-1}}{\tau_m}
\eeq

\noindent i.e., the equation of motion becomes a simpler drag equation for a
single body, given by 

\beq
\ddot{\v{r}}= -\frac{\mu\v{r}}{r^3} -\frac{\dot{\v{r}} - \v{u}_{\rm eff}}{\tau_{\rm eff}}-2n_{\rm out} \left(\hatz \times\dot{\v{r}}\right)+3n_{\rm out} ^2x\hatx -\frac{3n_{\rm out}  x \haty}{2\tau_{\rm eff}}
\label{eq:singleequiv}
\eeq

\noindent with effective wind 

\beq 
\v{u}_{\rm eff} = \v{u} \frac{\tau_m}{\tau_w} = \v{u} \frac{(\tau_2-\tau_1)(m_1+m_2)}{(m_2\tau_2+m_1\tau_1)}
\label{eq:equiv-wind}
\eeq

\noindent and effective drag time $\tau_{\rm eff} = \tau_m$. 

\subsection{A.2. Numerical validation of the single particle equivalent}

\begin{figure*}
  \begin{center}
    \resizebox{\textwidth}{!}{\includegraphics{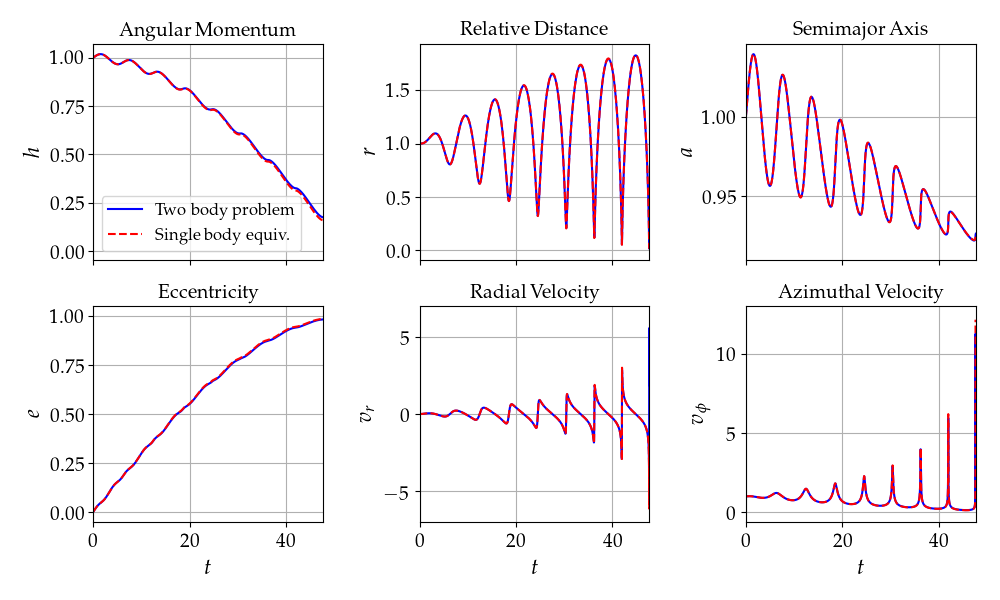}}
\end{center}
\caption{Comparison between the two massive bodies problem
  (\eq{eq:r1}-\eq{eq:r2}, solid blue line), and the equivalent 1-body
  system with reduced mass and effective wind and drag
  (\eq{eq:singleequiv}, dashed red line). The 2-body problem has $m_1=0.75$ and
  $m_2=0.25$, $\tau_1 = 3\times 10^3$ and $\tau_2 = 10^3$, and 
$u = 30$. The one-body equivalent has effective friction
time $\tau_{\rm eff} = 1200$, and effective
wind $u_{\rm eff} = 24$. The terms with $n_{\rm out} $ are
ignored. The agreement is excellent, validating the approximation
$u\gg \dot{s}_{\rm cm}$ that motivates \eq{eq:singleequiv}.}
\label{fig:equivalency}
\end{figure*}

We show in \fig{fig:equivalency} the evolution of the system \eq{eq:r1}-\eq{eq:r2}
and that of \eq{eq:singleequiv}. We model the system in a 2D Cartesian
box with a wind $\v{u} = u \hat{\v{y}}$, and ignoring $n_{\rm out}$. In code units we consider 

\beq
G = 1; \quad m_1+m_2=1; \quad a_0 = 1; \quad n_0 = 1. 
\eeq 

\noindent where $a_0$ is the initial semimajor axis and $n_0$ the
initial angular frequency of the binary. We solve the $N$-body with a standard Runge-Kutta scheme 3rd order accurate in
time. We take timesteps of
$\Delta t=10^{-3}T$, with the period $T=2\pi/n$ dynamically updated as the binary hardens. 

To test the code, we consider a binary system of arbitrary masses $m_1=0.75$ and $m_2=0.25$,
drag times $\tau_1 = 3\times 10^3$ and $\tau_2 = 10^3$, and a wind
$u = 30$. The masses' starting positions are given by \eq{eq:radii}
with $|\v{r}|=a$. The initial orbit is circular with velocities $v_{0i} = n_0 a_{0i}$. 

The simulation is centered at the center of mass and at
every full timestep the center of mass position and velocity are
reset. The numerical solution of this system, given by
\eq{eq:r1}-\eq{eq:r2}, is shown by the blue solid line
in \fig{fig:equivalency}. 

With these parameters, the one body equivalent \eqp{eq:singleequiv} has effective friction
time $\tau_{\rm eff} = 1200$ as given by \eq{eq:taueff}, and effective
wind $u_{\rm eff} = 24$, as given by \eq{eq:equiv-wind}. The numerical solution of this system, 
is shown by the red dashed line in \fig{fig:equivalency}. 

The only difference between these systems is that the single body
equivalent is missing the indirect term from the acceleration of the
center of mass. As evidenced by the similarity of the solutions, this
term is negligible, leading to but a minute deviation in angular
momentum and eccentricity toward contact (when the relative distance goes to zero). 

Finally, we notice that because the center of mass is accelerated, even
though initially the system may have $u\gg\dot{s}_{\rm cm}$, this assumption
may not be maintained during the course of the whole simulation. The
effect of the wind drag is to try to bring the center of mass velocity
to the same velocity as the wind, the situation where the wind
drag would cease to exist. The acceleration of the center of mass is
given by the center of mass equation 

\beqn
\ddot{\v{s}}_{\rm cm} &=& \frac{m_1 \ddot{\v{s}}_1   + m_2 \ddot{\v{s}}_2}{(m_1+m_2)}\nonumber\\
&=& \v{u} \frac{\tau_1m_2 + \tau_2m_1}{\tau_1\tau_2 (m_1+m_2)} -\dot{\v{s}}_1\frac{m_1}{(m_1+m_2)\tau_1}  - \dot{\v{s}}_2\frac{m_2}{(m_1+m_2)\tau_2}
\eeqn

If $u \gg \dot{s}_{1,2}$, the wind dominates; the center of mass will
accelerate, reaching velocity $u$ within the timescale $\tau_m$. 

\subsection{A.3. Orbital solution}

Since gravity dominates, we can treat the problem as a Keplerian orbit perturbed by the orbital
drag, the wind drag, the Coriolis force, the centrifugal force, and
the shear. We use the formalism of \cite{MurrayDermott99} to solve for the evolution of
semimajor axis, angular momentum, and eccentricity under these
forces. Treating the perturbation as 

\beq
d\v{F} = \overline{R}\hatr + \overline{T}\hatphi + \overline{N} \hath
\eeq

\noindent where $\hatr$, $\hatphi$ are the (cylindrical) unit vectors
in the plane of the orbit and $\phi=0$
points at pericenter; that is, $[\hatr,\hatphi]$ is
  $[\hate,\hatq]$ rotated by the true anomaly. 
The solutions for the orbital elements given by \cite[][eqs
2.145, 2.149, 2.150, and 2.157]{MurrayDermott99}, following \cite{Burns76}

\beqn
\frac{da}{dt} &=& \frac{2}{n\sqrt{1-e^2}} \ \chi_a \label{eq:adot}\\
\frac{dh}{dt} &=& \chi_h \label{eq:hdot}\\
\frac{de}{dt} &=& \frac{\sqrt{1-e^2}}{na} \ \chi_e \label{eq:edot}\\
\frac{dI}{dt} &=& \chi_I \label{eq:idot}\\
\frac{d\varOmega}{dt} &=& \chi_\varOmega \label{eq:Omegadot}\\
\frac{d\omega}{dt} &=& \frac{\sqrt{1-e^2}}{ane}\chi_\omega - \chi_\varOmega\cos I.\label{eq:omegadot}
\eeqn

\noindent with the functions $\chi$ given by 

\beqn
\chi_a           &=& \overline{R}e\sinf  + \overline{T}\left(1+e\cosf\right),\label{eq:chia}\\
\chi_h           &=& r\overline{T},\label{eq:chih}\\
\chi_e           &=& \overline{R}\sinf + \overline{T}\left(\cosf+\cosE\right),\label{eq:chie}\\
\chi_I            &=& r\overline{N} \ \frac{\cos(\omega+f)}{h},\label{eq:chiI}\\
\chi_\varOmega &=& r\overline{N} \ \frac{\sin(\omega+f)}{h\sin I},\label{eq:chiOmega}\\
\chi_\omega &=& -\overline{R}\cosf +  \overline{T}\sin f\left(\frac{2+ e \cos f}{1+e\cosf}\right), \label{eq:chiomega}
\eeqn

\noindent where $f$ is the true anomaly and $E$ is the eccentric
anomaly. We work out the functions $\overline{R}$, $\overline{T}$, and $\overline{N}$
for the several terms involved. 

\subsubsection{A.3.1. Orbital drag}

The orbital drag is 

\beq
\left[\begin{array}{c}
\overline{R}\\
\overline{T}\\
\overline{N}\end{array}\right]_{\rm drag} = 
-\frac{1}{\tau}\left[\begin{array}{c}
\dot{r}\\
r\dot{\phi}\\
0 \end{array}\right], \label{eq:dragRTN}
\eeq

\noindent for which we will need the solutions for $\dot{r}$  and $r\dot{\phi}$, 

\beqn
\dot{r}  &=& \frac{na}{\sqrt{1-e^2}}e\sinf,\label{eq:dotr}\\
r\dot{\phi}  &=& \frac{na}{\sqrt{1-e^2}}\left(1 + e \cosf\right).\label{eq:rdotphi}
\eeqn

These functions contain terms dependent on the true anomaly, so we
take orbital averages to find the secular evolution. We define orbital
averages as averages in mean anomaly $M=nt$ according to 

\beq
\mean{X} = \frac{1}{2\pi}\int_0^{2\pi} X \ dM 
\eeq

The series for $\sinf$ and $\cosf$ are, to 4th order in eccentricity,  

\beqn
\sinf &=&\sin M + e\sin 2M +e^2\left(\frac{9}{8}\sin 3M
  -\frac{7}{8}\sin M \right) +e^3\left(\frac{4}{3}\sin
  4M-\frac{7}{6}\sin 2M\right) \nonumber\\
&&+e^4\left(\frac{17}{192}\sin M -\frac{207}{128}\sin 3M +
  \frac{625}{384}\sin 5M\right) + \mathcal{O}(e^5)\label{eq:sinf}\\
\cosf &=& \cos M + e\left(\cos 2M -1\right) +\frac{9e^2}{8}\left(\cos
  3M +\cos M \right) +\frac{4e^3}{3}\left(\cos 4M-\cos 2M\right)
\nonumber\\
&&+e^4\left(\frac{25}{192}\cos M -\frac{225}{128}\cos 3M +
  \frac{625}{384}\cos 5M\right) + \mathcal{O}(e^5) \label{eq:cosf}
\eeqn

\noindent Clearly all terms
except $-e$ are periodic, so $\mean{\cosf} = -e$  and $\mean{\sinf}
=0$. The solution for $\cosE$ will also be needed, also shown to
fourth order in eccentricity

\beqn
\cosE &=& \cos M + \frac{e}{2}\left(\cos 2M -1\right)
+\frac{3e^2}{8}\left(\cos 3M -\cos M \right) +\frac{e^3}{3}\left(\cos
  4M-\cos 2M\right) \nonumber\\
&&+e^4\left(\frac{5}{192}\cos M -\frac{45}{128}\cos 3M + \frac{125}{384}\cos 5M\right) + \mathcal{O}(e^5)
\eeqn

\noindent All terms except $-e/2$ average out over an orbital period,
so $\mean{\cosE}=-e/2$. 

Next we write each perturbation term and find their effect on the
evolution of the orbital elements.

\paragraph{Semimajor axis}

Substituting \eq{eq:dotr} and \eq{eq:rdotphi} into \eq{eq:dragRTN},
and plugging into \eq{eq:chia} yields 

\beq
{\chi_a}_{, \rm drag} = -\frac{na}{\tau\sqrt{1-e^2}}\left(1+2e\cosf +e^2 \right).
\eeq

\noindent Now taking the orbital average using \eq{eq:cosf}, we find the contribution of
the orbital drag to the evolution of the semimajor axis 

\beq
\mean{\chi_a}_{\rm drag} = -\frac{na}{\tau}\sqrt{1-e^2}
\eeq

\paragraph{Angular momentum}

The angular momentum evolution is given by \eq{eq:hdot}, depending
only on the azimuthal part $\overline{T}$ of the
perturbation. Considering the drag 

\beq
\frac{dh}{dt} = - \frac{h}{\tau} 
\eeq

\paragraph{Eccentricity}

The evolution of eccentricity is given by \eq{eq:edot} and
\eq{eq:chie}. The effect of the drag on the eccentricity is, given \eq{eq:dragRTN} 

\beq
{\chi_e,}_{\rm drag} = -\frac{na}{\sqrt{1-e^2}}\frac{1}{\tau}\left(e + \cosf +\cosE  + e\cosE \cos f  \right)
\eeq

\noindent taking the average, $\mean{\cosf}=-e$ cancels with $e$. The average $\mean{\cosE} = -e/2$, so we have  

\beq
\mean{\chi_e}_{\rm drag} = -\frac{na}{\sqrt{1-e^2}}\frac{1}{\tau}\left(-e/2  + e\mean{\cosE \cos f}  \right)
\label{eq:edragmean}
\eeq

\noindent the average $\mean{\cosE \cosf}$ is found from the equation of the orbit 

\beq
r = a (1 - e \cosE ).
\eeq

\noindent Multiplying by $\cosf$ 

\beq
r\cosf = a(\cosf-e\cosE\cosf) 
\eeq

\noindent given $x_c=r\cosf$ and averaging 

\beq
\mean{x_c} = a(\mean{\cosf}-e\mean{\cosE\cosf}) 
\eeq

\noindent since $\mean{\cosf}=-e$ and $\mean{x_c} = -3 ae /2$, this
results in 

\beq 
\mean{\cosE\cosf}=\frac{1}{2}.
\eeq

The term in parentheses in \eq{eq:edragmean} thus cancels out exactly,
so $\mean{\chi_e}_{\rm drag}$ = 0 and the orbital drag does not affect the eccentricity. \\

\paragraph{Inclination}

The evolution of inclination is given by \eq{eq:idot}. The orbital
drag does not have a normal component, so it cannot affect the
inclination. 

\paragraph{Longitude of ascending node}

The expression for the evolution of the longitude of the ascending
node is similar to the one for the inclination. The orbital drag does
not have a normal component and thus has
no effect. 

\paragraph{Argument of Pericenter }

The evolution of the argument of pericenter is given by
\eq{eq:omegadot} and \eq{eq:chiomega}. For the orbital drag, the contribution is

\beq
{\chi_\omega,}_{\rm drag} = \frac{ 2 a n \sinf } {\tau\sqrt{1-e^2}} 
\eeq

\noindent which integrates to zero. The orbital drag does not lead
to precession. 

\paragraph{Orbital Drag: summary}

The orbital drag contribution is

\beqn
\mean{\chi_a}_{\rm drag}  &=& -\frac{na}{\tau}\sqrt{1-e^2}\\
\mean{\chi_h}_{\rm drag} &=& - \frac{h}{\tau} \\
\mean{\chi_e}_{\rm drag}  &=&  0\\
\mean{\chi_I}_{\rm drag}  &=& 0 \\
\mean{\chi_\varOmega}_{\rm drag} &=& 0 \\
\mean{\chi_\omega}_{\rm drag}  &=& 0
\eeqn

\subsubsection{A.3.2. Wind}

As for the wind, it is always blowing from the $\haty$ direction in
Hill Cartesian coordinates $\v{x}$. We transform between these
coordinates and the (Cartesian) orbital plane coordinates $\v{x}_{\rm
  cart}=x_c \hate + y_c \hatq + z_c \hath$ according to 

\beq
\v{x}_{\rm cart} = \vt{R}_h(\omega) \vt{R}_e(I) \vt{R}_h(\varOmega)\v{x}
\eeq

Where $\vt{R}_j$ is the rotation matrix about axis $j$. To pass to the
orbital plane in cylindrical coordinates $\hatr \hatphi \hath$, we
rotate clockwise around $\hath$ by the true anomaly, i.e.,
$\v{x}_{\rm cyl} = \vt{R}_h(-f)\v{x}_{\rm cart}$. We thus have 

\beq
 \v{x}_{\rm cyl} = \vt{R}\v{x} =  \vt{R}_h(\omega-f) \vt{R}_e(I) \vt{R}_h(\varOmega)\v{x}
\eeq

\noindent where $\vt{R} =   \vt{R}_h(\omega-f) \vt{R}_e(I) \vt{R}_h(\varOmega)$ is the
full rotation matrix for the transformation. For the wind, the vector is in the $\haty$ direction; for completeness
we give the transformations of the local Hill coordinate unit vectors $\hatx=[1,0,0]^T$,
$\haty=[0,1,0]^T$, and $\hatz=[0,0,1]^T$ to the coordinate system 
$\hatr\hatphi\hath$ of the binary orbit 

\beq
R \hatx = \left[\begin{array}{c}
\cos I\sin \varOmega\sinfw +\cosfw\cos\varOmega\\
-\cos\varOmega\sinfw +\cos I\sin \varOmega\cosfw\\
\sin\varOmega\sin I\end{array}\right],
\eeq

\beq 
R \haty = \left[\begin{array}{c}
\cos I\cos \varOmega\sinfw -\cosfw\sin\varOmega\\
\sin\varOmega\sinfw +\cos I\cos \varOmega\cosfw\\
\cos\varOmega\sin I\end{array}\right],
\eeq
\noindent and
\beq 
R \hatz = \left[\begin{array}{c}
-\sin I\sinfw\\
-\sin I\cosfw\\
\cos I\end{array}\right].
\label{eq:rotxyz}
\eeq

Thus, for the wind 

\beq
\left[\begin{array}{c}
\overline{R}\\
\overline{T}\\
\overline{N}\end{array}\right]_{\rm wind} = 
-\frac{u}{\tau}\left[\begin{array}{c}
\cos I\cos \varOmega\sinfw -\cosfw\sin\varOmega\\
\sin\varOmega\sinfw +\cos I\cos \varOmega\cosfw\\
\cos\varOmega\sin I\end{array}\right]. \label{eq:windRTN}
\eeq

\paragraph{Semimajor axis}

The influence on the semimajor axis, given by \eq{eq:chia}, is 

\beq
\chi_{a,_{\rm wind}} = \cos I \cos\varOmega \left[ \cosfw + e\cos\omega \right] + \sin\varOmega\left[\sinfw -e\sin\omega\right]
\label{eq:chiawind}
\eeq

Taking the orbital average,  

\beqn 
\mean{\cosfw} &=& \mean{\cos  f}\cos\omega + \mean{\sin f}\cos\omega =
-e\cos\omega, \\
\mean{\sinfw} &=& \mean{\sin f}\cos\omega -\mean{\cos f}\sin\omega =
e\sin\omega. 
\eeqn

Averaged in the inner orbit, all terms in \eq{eq:chiawind} cancel, i.e. 

\beq
\mean{\chi_a}_{\rm wind} =0.
\eeq

\noindent The external wind has no secular effect on the semimajor axis. 

\paragraph{Angular momentum}

For the wind

\beqn
\frac{dh}{dt} &=& - \frac{u}{\tau} r \left[\sin\varOmega  \sinfw +\cos I\cos \varOmega \cosfw\right]\\
&=& - \frac{u}{\tau} \left[\sin\varOmega (y\cos\omega - x \sin\omega) +\cos I\cos \varOmega (x\cos\omega + y\sin\omega)\right]
\eeqn

\noindent where $x=r\cosf$ and $y=r\sinf$ are the Cartesian 
coordinates in the reference frame of the orbit. Given the
Keplerian solution, they are $x=a(\cos E - e)$ and
$y=a\sin E$. Using the expansion for $E$, it results in
$\mean{x} = -3ae/2$ and $\mean{y}=0.$ So, 

\beq
\frac{d\mean{h}}{dt} = - ae\frac{3u}{2\tau}\left(\cos I\cos \varOmega \cos\omega - \sin\varOmega \sin\omega \right)
\eeq

\paragraph{Eccentricity}

For the wind, according to \eq{eq:windRTN}

\beq
{\chi_e}_{, {\rm wind}}  = \frac{u}{\tau} \left\{  \cos I \cos\varOmega \left[\cos E\cosfw + \cos\omega \right]   +  \sin\varOmega \left[ \cos E \sinfw -\sin\omega\right]\right\}
\eeq

 given $\mean{\cosE \cosf}=1/2$ and  $\mean{\cosE \sinf}=0$, the
average over $f$ is 

\beq
\mean{\chi_e}_{\rm wind}  = \frac{3u}{2\tau} \left(  \cos I \cos\varOmega  \cos\omega   -  \sin\varOmega \sin\omega \right)
\eeq

The evolution of the orbitally-averaged eccentricity is thus due to
the wind only, according to 

\beq
\frac{d\mean{e}}{dt} = \frac{\sqrt{1-e^2}}{na} \ \frac{3u}{2\tau} \ \left(  \cos I \cos\varOmega  \cos\omega   -  \sin\varOmega \sin\omega \right)
\eeq

\paragraph{Inclination}

The evolution of inclination is given by \eq{eq:idot}. For the wind  

\beq
\frac{dI}{dt} =  -\frac{u}{h\tau} a\cos\varOmega\sin I \left[ \cos(f+\omega)\right] (1 - e\cos E)
\eeq

This expression expands to 

\beq
\frac{dI}{dt} =  -\frac{u}{h\tau} a\cos\varOmega\sin I 
\left(
  \cos\omega\cosf - \sin\omega\sinf - e\cos\omega \cos E\cos f + e  \sin\omega \cos E\sin f\right) 
\eeq

On averaging, the second and last terms in parentheses cancel out. The
first and third terms add up to $-3/2ae \cos\omega$ . The evolution
of the orbit-averaged inclination due to the wind is thus 

\beq
\frac{d\mean{I}}{dt} =  \frac{3u}{2\tau} \frac{ae}{h} \cos\varOmega\cos \omega \sin I
\eeq

\paragraph{Longitude of ascending node}

The effect of the wind is 

\beq
\frac{d\varOmega}{dt} =  -\frac{u}{h\tau}  a\cos\varOmega\sin(f+\omega) ( 1 - e \cos E) 
\eeq

\noindent which expands to 

\beq
\frac{d\varOmega}{dt} =  -\frac{u}{h\tau}  a\cos\varOmega\left(
  \cos\omega\sin f + \sin \omega \cos f - e\cos\omega \cos E\sin f - e
  \sin\omega \cos E\cos f\right) 
\eeq

On averaging, the first and third terms in parentheses cancel out. The
second and last terms add up to $-3/2ae \sin\omega$. The evolution
of the orbit-averaged longitude of ascending node is thus 

\beq
\frac{d\mean{\varOmega}}{dt} =  \frac{3u}{2\tau}  \frac{ae}{h} \cos\varOmega\sin\omega.
\eeq

\paragraph{Argument of Pericenter}

The evolution of the argument of pericenter is given by
\eq{eq:omegadot} and \eq{eq:chiomega}. For the wind

\beqn
\chi_\omega &=& \frac{u}{\tau}\left\{\cosf\left[\cos
    I\cos\varOmega\sinfw-\cosfw\sin\varOmega\right]\right. \nonumber \\
&&\left.-\left(\frac{2+e\cosf}{1+e\cosf}\right)\sinf\left[\sin\varOmega\sinfw+\cos I\cos\varOmega\cosfw\right]\right\}
\eeqn

\noindent we expand and group the terms as 

\beqn
\chi_\omega &=& \frac{u}{\tau}\left\{  -(\cosI\cosO\sinw+\cosw\sinO) \left[\cos^2f +\sin^2f \left(\frac{2+e\cosf}{1+e\cosf}\right)\right]\right.\nonumber\\
&&\left.+(\cosI\cosO\cosw-\sinO\sinw)\left[1 -\left(\frac{2+e\cosf}{1+e\cosf}\right)\right]\cosf\sinf  \right\}
\eeqn

\noindent upon integration the second term is periodic in $M$ and cancels. We
are left with 

\beq
\chi_\omega = \frac{u}{\tau} \left(\cosI\cosO\sinw+\cosw\sinO\right)  A(e) 
\eeq 

\noindent where 

\beq
A(e) = -\mean{\cos^2f} - \mean{\sin^2f\left(\frac{2+e\cosf}{1+e\cosf}\right)}
\eeq

\noindent a function of the eccentricity alone. The orbital evolution of the argument of pericenter is thus 

\beq
\frac{d\mean{\omega}}{dt} =  \frac{u}{\tau}\frac{\sqrt{1-e^2}}{ane} \left\{ \left[A(e)  - \frac{3}{2}  \frac{ae}{h}\right]\cos I \cos\varOmega \sin\omega + A(e)\cos\omega\sin\varOmega \right\}
\eeq

\paragraph{Wind: summary}

The external wind contribution is

\beqn
\mean{\chi_a}_{\rm wind}  &=& 0\\
\mean{\chi_h}_{\rm wind} &=& - ae\frac{3u}{2\tau}\left(\cos I\cos \varOmega \cos\omega - \sin\varOmega \sin\omega \right)\\
\mean{\chi_e}_{\rm wind}  &=& \frac{\sqrt{1-e^2}}{na} \ \frac{3u}{2\tau} \ \left(  \cos I \cos\varOmega  \cos\omega   -  \sin\varOmega \sin\omega \right)\\
\mean{\chi_I}_{\rm wind}  &=& \frac{3u}{2\tau} \frac{ae}{h} \cos\varOmega\cos \omega \sin I\\
\mean{\chi_\varOmega}_{\rm wind} &=& \frac{3u}{2\tau}  \frac{ae}{h} \cos\varOmega\sin\omega\\
\mean{\chi_\omega}_{\rm wind}  &=&\frac{u}{\tau} \left(\cosI\cosO\sinw+\cosw\sinO\right)  A(e) 
\eeqn

\subsubsection{A.3.3. Coriolis force}

The Coriolis force, being an inertial force, cannot alter the
energy or angular momentum of the orbit. As a consequence,
eccentricity is also unmodified. Its effect is to
lead to an apparent precession of the orbit in the Hill co-rotating coordinate
frame. We work out the perturbations introduced by the Coriolis Given Eqs.~\ref{eq:rotxyz}

\beq
\left[\begin{array}{c}
\overline{R}\\
\overline{T}\\
\overline{N}\end{array}\right]_{\rm Coriolis} = 
2n_{\rm out}  \left[\begin{array}{c}
		r\dot{\phi} \cos I \\
		-\dot{r}\cos I\\
		r\dot{\phi}\sin I\sinfw  - \dot{r}\sin I \cosfw 
\end{array}\right]
\eeq

\paragraph{Semimajor axis}

For the semimajor axis, using \eq{eq:chia}, 

\beq
{\chi_h,}_{\rm Coriolis} = 2n_{\rm out} \cos I \left[  r\dot{\phi} e\sinf  -\dot{r}\left(1+e\cosf\right)    \right]
\eeq

\noindent and given \eq{eq:dotr} and \eq{eq:rdotphi} the two terms
cancel identically: $\chi_a=0$.

\paragraph{Angular momentum}

The influence of the Coriolis force on angular momentum is 

\beq
{\chi_h,}_{\rm Coriolis} = -\frac{2n_{\rm out}  n a^2e \cos I }{\sqrt{1-e^2}} \left(1-e\cosE\right) \sinf 
\eeq

\noindent which is a periodic function of $M$ and integrates to zero. 

\paragraph{Eccentricity}

The influence of the Coriolis force on eccentricity is 

\beq
{\chi_e,}_{\rm Coriolis} = \frac{2n_{\rm out}  n a \cos I }{\sqrt{1-e^2}}\left( 1-e\cosE\right) \sinf          
\eeq

\noindent which is a periodic function of $M$ and integrates to zero. 

\paragraph{Inclination}

The influence of the Coriolis force on inclination is 

\beq
{\chi_I,}_{\rm Coriolis} = -\frac{2 n_{\rm out}  n a^2 \sin I}{\sqrt{1-e^2}}   [e\sin\omega - \sin(f-\omega)] \cos(f+\omega)  (1-e\cosE)
\eeq

The product $\sin(f-\omega) \cos(f+\omega)  =  \cosf\sinf - \cos\omega\sin\omega$, so 

\beq
{\chi_I,}_{\rm Coriolis} = -\frac{2 n_{\rm out}  n a^2 \sin I}{\sqrt{1-e^2}}\psi_I 
\eeq

\noindent where

\beq
\psi_I = [(1+e\cosf) \sin\omega\cos\omega-e\sin\omega^2\sinf -\cosf\sinf]  (1-e\cosE).
\eeq

\noindent And, expanding these terms, 

\beqn
\psi_I &=& (1+e\cosf - e\cosE -e^2\cosf\cosE) \sin\omega\cos\omega\nonumber\\
&&-e\sin\omega^2\sinf -\cosf\sinf  + e2\sin\omega^2\cosE\sinf -e\cosE\cosf\sinf
\eeqn

Integrating, all terms but the first one cancel out, leaving only 

\beq
\mean{\psi_I} = (1-e^2) \sin\omega\cos\omega
\eeq

The evolution of inclination due to the Coriolis force is thus 

\noindent 

\beq
{\chi_I,}_{\rm Coriolis} = -2 n_{\rm out}  n a^2 \sqrt{1-e^2} \sin I \sin\omega\cos\omega
\eeq

\paragraph{Longitude of ascending node }

The influence of the Coriolis force on the longitude of the ascending node is  

\beq
{\chi_\varOmega,}_{\rm Coriolis} = -\frac{2 n_{\rm out}  n a^2}{\sqrt{1-e^2}}   [e\sin\omega - \sin(f-\omega)] \sin(f+\omega)  (1-e\cosE)
\eeq

The product $\sin(f-\omega) \sin(f+\omega)  =  -1/2 \cos 2f + 1/2 \cos 2\omega$, so 

\beq
{\chi_\varOmega,}_{\rm Coriolis} = -\frac{2 n_{\rm out}  n a^2}{\sqrt{1-e^2}}\psi_\varOmega 
\eeq

\noindent where $\psi_\varOmega = (e\sinf\cos\omega\sin\omega + e\sin^2\omega\cosf + 1/2\cos 2f - 1/2 \cos 2\omega)  (1-e\cosE)$

\noindent expanding these terms, 

\beqn
\psi_\varOmega &=& e\sin\omega\cos\omega\sinf+ e\sin^2 \omega\cosf+ 1/2 \cos 2f - 1/2 \cos 2\omega\nonumber\\
&&-e^2\sin \omega\cos \omega  \sinf\cosE-e^2\sin^2 \omega\cosf\cosE-e\cosE \cos 2f / 2+ e\cosE \cos 2\omega / 2
\eeqn

Integrating, we are left with 

\beq
\mean{\psi_\varOmega} = \frac{1}{2} \left[- 3 e^2\sin 2\omega-\left(1+\frac{e^2}{2}\right) \cos 2\omega + \mean{\frac{r}{a}  \cos2f}\right]
\eeq

\noindent writing 

\beq
B(e) = \mean{\frac{r}{a}  \cos2f}
\eeq

\noindent the evolution of longitude of ascending node due to the Coriolis force is 

\beq
{\chi_\varOmega,}_{\rm Coriolis} = -\frac{n_{\rm out}  n a^2}{\sqrt{1-e^2}} \left[- 3 e^2\sin 2\omega-\left(1+\frac{e^2}{2}\right) \cos 2\omega + B(e)\right]
\eeq

\paragraph{Argument of Pericenter}

The evolution of the argument of pericenter is given by 

\beq
{\chi_\omega,}_{\rm Coriolis} = -\frac{2  n_{\rm out}   n a\cos I  }{\sqrt{1-e^2}} \psi_\omega
\eeq

\noindent with 

\beq
\psi_\omega = \cosf + e\cos^2f + e \sin^2f\left(\frac{2+e\cosf}{1+e\cosf}\right) 
\eeq

\noindent upon integration 

\beq
\mean{\psi_\omega} = -e + e\mean{\cos^2f} + e \mean{\sin^2f\left(\frac{2+e\cosf}{1+e\cosf}\right) }
\eeq

a function of the eccentricity alone. The evolution of the argument of
pericenter is thus 

\beq
{\chi_\omega,}_{\rm Coriolis} = -\frac{2  n_{\rm out}   n a\cos I
}{\sqrt{1-e^2}} C(e) 
\eeq

where $C(e) = \mean{\psi_\omega}$. 

\paragraph{Coriolis force: summary}

The Coriolis force contribution is

\beqn
\mean{\chi_a}_{\rm Coriolis}  &=& 0\\
\mean{\chi_h}_{\rm Coriolis} &=& 0\\
\mean{\chi_e}_{\rm Coriolis}  &=& 0\\
\mean{\chi_I}_{\rm Coriolis}  &=& -2 n_{\rm out}  n a^2 \sqrt{1-e^2} \sin I \sin\omega\cos\omega\\
\mean{\chi_\varOmega}_{\rm Coriolis} &=& -\frac{     n_{\rm out}   n a^2}{\sqrt{1-e^2}} \left[- 3 e^2\sin 2\omega-\left(1+\frac{e^2}{2}\right) \cos 2\omega + B(e)\right]\\
\mean{\chi_\omega}_{\rm Coriolis}  &=& -\frac{2  n_{\rm out}   n a\cos I}{\sqrt{1-e^2}} C(e) 
\eeqn

\subsection{A.4. Orbital Evolution}

Putting it all together (ignoring centrifugal force and shear) 

\beqn
\frac{d\mean{a}}{dt} &=& -\frac{2\mean{a}}{\tau};\\
\frac{d\mean{e}}{dt} &=& \frac{\sqrt{1-e^2}}{na}\frac{3u}{2\tau}\left(  \cos I \cos\varOmega  \cos\omega   -  \sin\varOmega \sin\omega \right);\\
\frac{d\mean{h}}{dt}&=& -\frac{\mean{h}}{\tau} - ae\frac{3u}{2\tau} \left(\cos I\cos \varOmega \cos\omega-\sin\varOmega \sin\omega\right);\\
\frac{d\mean{I}}{dt} &=&  \frac{3u}{2\tau} \frac{ae}{h} \cos\varOmega\cos\omega \sin I -2 n_{\rm out}  n a^2 \sqrt{1-e^2} \sin I \sin\omega\cos\omega;\\
\frac{d\mean{\varOmega}}{dt} &=&  \frac{3u}{2\tau}  \frac{ae}{h} \cos\varOmega\sin\omega -\frac{n_{\rm out}  n a^2}{\sqrt{1-e^2}} \left[- 3 e^2\sin 2\omega-\left(1+\frac{e^2}{2}\right) \cos 2\omega + B(e)\right];\\
\frac{d\mean{\omega}}{dt} &=&  \frac{\sqrt{1-e^2}}{ane} \left\{\frac{u}{\tau}\left[A(e)  - \frac{3}{2}  \frac{ae}{h}\right]\left[\cos I \cos\varOmega \sin\omega + A(e) \cos\omega\sin\varOmega\right]\right. \\\nonumber 
&&\left. -\frac{2  n_{\rm out}   n a\cos I}{\sqrt{1-e^2}} C(e) + \frac{2n_{\rm out}  n a^2\cos I }{\sqrt{1-e^2}} \left[- \frac{3}{2} e^2\sin 2\omega-\frac{1}{2}\left(1+\frac{e^2}{2}\right) \cos 2\omega + \frac{B(e)}{2}\right]\right\}.
\eeqn

The system is over-specified because $a$ and $h$ define the
eccentricity. Still we keep the equation for $h$ for physical
insight. 

\subsubsection{A.4.1. Isolated binary ($n_{\rm out} =0$)}

Let us consider first the case where $n_{\rm out} =0$, i.e., an isolated
binary not in orbit around the Sun. The equations reduce to 

\beqn
\frac{d\mean{a}}{dt} &=& -\frac{2\mean{a}}{\tau};\\
\frac{d\mean{e}}{dt} &=& \frac{\sqrt{1-e^2}}{na}\frac{3u}{2\tau}\left(  \cos I \cos\varOmega  \cos\omega   -  \sin\varOmega \sin\omega \right);\\
\frac{d\mean{h}}{dt}&=& -\frac{\mean{h}}{\tau} - ae\frac{3u}{2\tau} \left(\cos I\cos \varOmega \cos\omega-\sin\varOmega \sin\omega\right);\\
\frac{d\mean{I}}{dt} &=&  \frac{3u}{2\tau} \frac{ae}{h} \cos\varOmega\cos \omega \sin I;\\
\frac{d\mean{\varOmega}}{dt} &=&  \frac{3u}{2\tau}  \frac{ae}{h} \cos\varOmega\sin\omega;\\
\frac{d\mean{\omega}}{dt} &=&  \frac{\sqrt{1-e^2}}{ane} \left[ \left(A(e)  - \frac{3u}{2\tau}  \frac{ae}{h}\right)\cos I \cos\varOmega \sin\omega + A(e) \cos\omega\sin\varOmega \right].
\eeqn

For an orbit originally at $\varOmega=\omega=0$, the
derivatives of $\mean{\varOmega}$, and $\mean{\omega}$ vanish. This is
a remarkable effect: there is no precession of the argument of
pericenter or longitude of the ascending node for this choice of
parameters. The system reduces to 

\beqn
\frac{d\mean{a}}{dt} &=& -\frac{2\mean{a}}{\tau};\label{eq:ared}\\
\frac{d\mean{e}}{dt} &=&\frac{\sqrt{1-e^2}}{na}\frac{3u}{2\tau}\cos I;\\
\frac{d\mean{h}}{dt}&=& -\frac{\mean{h}}{\tau} - ae\frac{3u}{2\tau}\cos I; \\
\frac{d\mean{I}}{dt} &=&  \frac{3u}{2\tau} \frac{ae}{h} \sin I;\label{eq:Ired}
\eeqn

\noindent which is not decoupled because eccentricity and inclination
depend on each other. For zero initial inclination the derivative of
$\mean{I}$ also cancels, and the system further reduces to

\beqn
\frac{d\mean{a}}{dt} &=& -\frac{2\mean{a}}{\tau};\\
\frac{d\mean{e}}{dt} &=& \frac{\sqrt{1-e^2}}{na}\frac{3u}{2\tau};\\
\frac{d\mean{h}}{dt}&=& -\frac{\mean{h}}{\tau} - ae \frac{3u}{2\tau}. 
\eeqn

\begin{figure*}
  \begin{center}
    \resizebox{\textwidth}{!}{\includegraphics{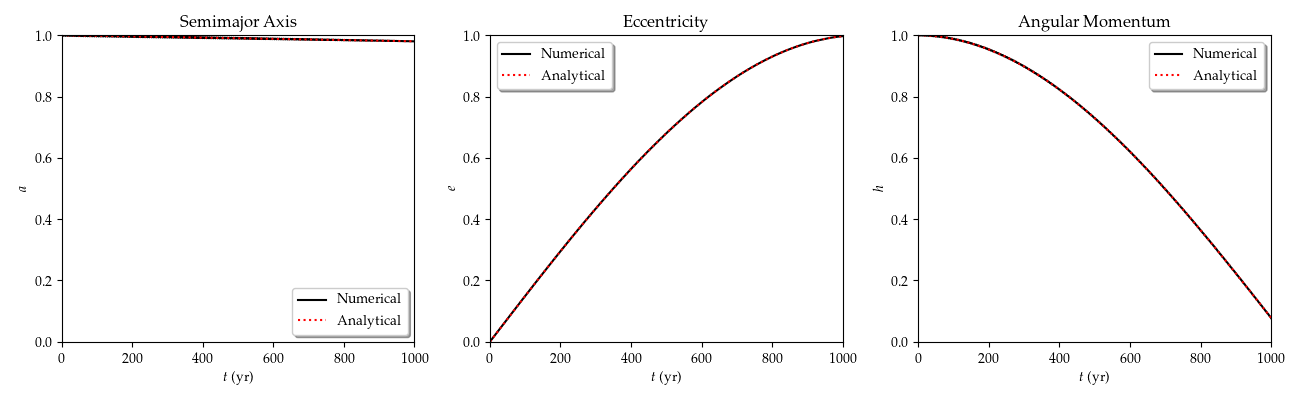}}
\end{center}
\caption{The evolution of an isolated binary under influence of wind
  and orbital drag. The wind
  drives angular momentum loss while it has no effect
  on the energy (which is dissipated via orbital drag). The effect is
  rapid eccentricity growth at nearly constant semimajor axis. Contact
  happens when the eccentricity nears one. The numerical calculation is
  in excellent agreement with the analytical solution \eqss{eq:asol}{eq:hsol}.}
\label{fig:IsolatedBinary}
\end{figure*}

This is a system that loses energy in a slow timescale given by $\tau/2$, whereas the angular
momentum decreases (thus eccentricity increases) in the faster timescale
given by the wind. The general solution for  $I=\varOmega=\omega=0$ is 

\beqn
\mean{a(t)} &=& a_0 \ee^{-2t/\tau};\label{eq:asol}\\
\mean{e(t)} &=& \cos\left[\cos^{-1}\left(e_0\right)  + \frac{3u}{2}\sqrt{\frac{a_0}{\mu}}\left(1-\ee^{-t/\tau}\right)\right];\label{eq:esol}\\
\mean{h(t)} &=& \ee^{-t/\tau}\left\{h_0-1+\cos\left[ \frac{3}{2}a_0 u\left(1-\ee^{-t/\tau}\right)     \right] \right\};\label{eq:hsol}
\eeqn

\noindent which we show graphically in \fig{fig:IsolatedBinary}, the agreement
between the analytical and numerical solutions is excellent.

\subsubsection{A.4.2. Hierarchical binary ($n_{\rm out} \neq 0$)}

For an orbit originally at $\varOmega=\omega=0$, the
derivatives of $\mean{\varOmega}$, and $\mean{\omega}$ are 

\beqn
\frac{d\mean{\varOmega}}{dt} &=&  \frac{n_{\rm out} n a^2}{\sqrt{1-e^2}} D(e);\\
\frac{d\mean{\omega}}{dt} &=&   n_{\rm out} F(e)\cos I .
\eeqn

\beqn
D(e) &=& \left[\left(1+\frac{e^2}{2}\right) - B(e)\right]\\
F(e) &=& \frac{1}{e}\left[-\left(1+\frac{e^2}{2}\right)  +  B(e)\right] + C(e) 
\eeqn

\noindent if $e$ and $I$ are slow-growing, then $\omega = F(e) \cos I
\ n_{\rm out} t  $. The mean anomaly of the
  outer orbit $M_{\rm out} = n_{\rm out} t$  is thus related to the argument of pericenter.

\beq
\omega = F(e) \cos I \ M_{\rm out}
\eeq

Thus, if eccentricity and inclination are slow growing in comparison
to $M_{\rm out}$ we can approximate 

\beq
d\omega \approx F(e) \cos I \ dM_{\rm out}
\eeq

\noindent if we define an average over the solar period, 

\beq
\tilde{X}  = \frac{1}{2\pi}\int_0^{2\pi} X dM_{\rm out}
\eeq

This average can be related to an average in argument of pericenter 

\beq
\tilde{X}  = \frac{1}{2\pi F(e)\cos I }\int_0^{2\pi} X d\omega
\eeq

\begin{figure*}
  \begin{center}
    \resizebox{\textwidth}{!}{\includegraphics{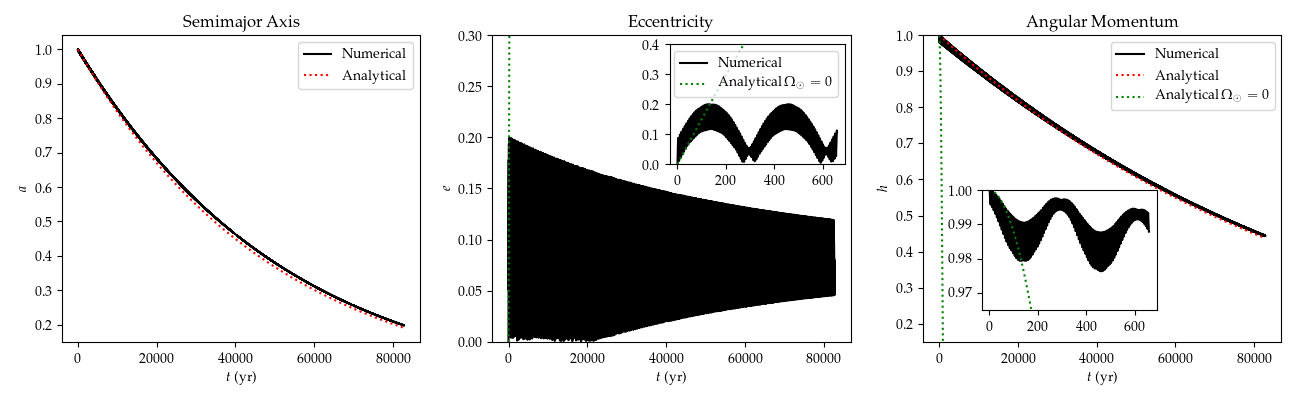}}
\end{center}
\caption{Evolution considering the orbit around the Sun. As the
  direction of the wind changes with respected to the fixed
  eccentricity vector of the binary, for half of the solar orbit the
  eccentricity grows as predicted by the isolated binary solution, and
  for the other half it decreases by the same amount. Similarly, the
  angular momentum decreases by the wind, and then increases in the
  other half of the orbit. Averaged over the solar period, the
  solution is described by a simple orbital drag solution 
  $a =a_0\ee^{-2t/\tau}$, $e\equiv {\rm const}$ and $h = h_0\ee^{-t/\tau}$.} 
\label{fig:Binary}
\end{figure*}

Thus, averaging over a precession period (related to the solar
period), the equations for the other parameters reduce to 

\beqn
\frac{d\tilde{a}}{dt} &=& -\frac{2\tilde{a}}{\tau};\\
\frac{d\tilde{e}}{dt} &=& 0;\\
\frac{d\tilde{h}}{dt}&=& -\frac{\tilde{h}}{\tau};\\
\frac{d\tilde{I}}{dt} &=&  0.
\eeqn

\noindent the eccentricity and inclination variation cancel out, as well as the wind term in the angular
momentum evolution.  During a solar orbit period, the wind makes the
eccentricity grow and angular momentum decay for
half the orbit, and then decrease by the same amount in the other
half. Energy and angular momentum decay at the timescale of orbital
drag $\tau$ while keeping the eccentricity constant. This
behavior is shown in \fig{fig:Binary}. Averaged over orbital and solar period,
only the orbital drag remains and the solution is simply 

\beqn
\tilde{a} &=& a_0\ee^{-2t/\tau};\\
\tilde{e} &=& e_0;\\
\tilde{h} &=& h_0\ee^{-t/\tau};\\
\tilde{I} &=& I_0.
\eeqn

\section{B. Drag Time}
\label{app:dragtime}

When the mean free path of the gas is much smaller than the object, the gas can be treated
like a fluid and viscous interactions at the surface of the body lead to
the emergence of drag. The mean free path is 

\beq
\lambda_{\rm mfp} = \frac{\mu_{\rm mol} m_H}{\rho  \sigma_{\rm coll}}
\eeq

\noindent where $\sigma_{\rm coll} = 2\times 10^{-15} \,{\rm cm}^2$ is
the collisional cross section of molecular hydrogen, $\mu_{\rm
  mol}=2.3$ is the mean molecular weight for a 5:2 hydrogen to
helium mixture, $\rho$ is the gas volume density, and $m_H$ stands for the atomic mass unit. 

Using the MMSN temperature and column density \citep{Weidenschilling77,Hayashi81,ChiangGoldreich97}

\beqn
{\rm T} &=& 280\,{\rm K} \  \left(\frac{r}{1\,{\rm AU}}\right)^{-0.5} \\ 
\Sigma &=& 1700 \, {\rm g\,cm}^{-2}  \left(\frac{r}{1\,{\rm AU}}\right)^{-1.5},
\eeqn

\noindent one finds $\lambda_{\rm mfp} =0.5$\,km at 45\,AU in the
MMSN, and the drag regime of continuous flow is valid. This regime 
splits into two regimes depending on the Reynolds
number, linear and quadratic, with
a smooth transition in between. Stokes drag
happens for small Reynolds number ($\Rey \lesssim 1$ ), where the drag is
dominated by viscosity at the surface of the body. The transition to
quadratic drag happens at  high Reynolds
numbers ($\Rey \gtrsim 800$), where ram pressure dominates. The
Reynolds number is 

\beq
\Rey=2R \rho |\Delta \v{v}|/\mu_{\rm visc},
\eeq

\noindent where $\Delta  \v{v}$ is the relative velocity between the
body and the gas and

\beq
\mu_{\rm visc}=\sqrt{\frac{8}{9\pi}} \rho c_s \lambda_{\rm mfp}
\eeq

\noindent is the dynamical viscosity, with $c_s$ being the sound speed. Substituting this expression
into $\Rey$, with $\Delta v = \eta v_k$ for the wind, leads to 

\beqn
\Rey &=& \frac{3}{4}\left\vert\frac{\partial\ln P}{\partial\ln r}\right\vert\frac{\sigma_{\rm coll}}{m_H} \frac{R}{\mu_{\rm
          mol}}      \frac{\Sigma}{r}\\
&\approx& 3 \left\vert\frac{\partial\ln P}{\partial\ln r}\right\vert\left(\frac{R}{10\,{\rm
    km}}\right)\left(\frac{\mu_{\rm mol}}{2}\right)^{-1}\left(\frac{\varSigma}{5\,{\rm
    g\,cm^{-2}}}\right)\left(\frac{r}{45\,{\rm AU}}\right)^{-1-p} \nonumber 
\eeqn

\noindent where $p\equiv -\partial\ln \Sigma / \partial\ln r$  is the power law of the column
density, positively defined. For the MMSN, $\Rey \approx 10$ at
45\,AU, and thus we are very close to Stokes law. The drag time is 

\beq
\tau = \frac{4\lambda_{\rm mfp}\rho_\bullet}{3\rho C_D c_s}  \frac{1}{\Ma\Kn}
\eeq

\noindent where $\Kn = \lambda_{\rm mfp}/2R$ is the Knudsen number and
$\Ma=|\Delta \v{v}|/c_s$ the flow Mach number. For Stokes flow at low Reynolds number
$C_D=24/\Rey$, leading to 

\beq
\tau = \frac{16}{18}  \frac{\rho_\bullet  R^2}{c_s}\frac{\sigma_{\rm coll}}{\mu  m_H}.
\eeq

The resulting drag times are $\tau_1=4.72\times 10^7$\,yr and $\tau_2 = 3.13 \times 10^7$
yr for the pre-merger lobes of MU69 in the low Reynolds number
regime. For arbitrary Reynolds number the drag coefficient $C_D$ is
given by \eq{eq:cd}, leading to the values of $\tau_1=2.87\times 10^7$\,yr and $\tau_2 = 2.00 \times
10^7$\,yr quoted in \sect{sect:drag}.

\section{C. Single vs double averaged secular dynamics} 
\label{app:single-double}

Here we consider the applicability of secular dynamics, specifically Kozai-Lidov oscillations, for KBO binaries.  The standard formulae for Kozai oscillations occur in the double-averaged (in time) approximation, taken to quadrupole order (in distance).  For KBOs the binary separation $a$ is much less than the distance to the Sun $a_{\rm out} \simeq 44.5 $ AU (numerical value for MU69 adopted).  Thus the quadrupole approximation should be more than sufficiently accurate.

As for the double-averaged approximation we consider the criterion given in \citep[see their Eq.\ 20, and references therein]{Liu+19} which states that  the eccentricity change timescale should be longer than the outer period (thus making is appropriate to take the secular average over the outer orbit):
\begin{align}\label{eq:doubleav}
t_{\rm kozai} \sqrt{1 - e^2} & \gtrsim T_{\rm out}
\end{align}

\noindent where no subscript refers to the inner
  binary and the subscript `out' refers to the outer (object 3) orbit.

\begin{align}
  t_{\rm kozai} &= n^{-1} \frac{(m_1+m_2)}{m_3} \left(\frac{a_{\rm out}}{a} \right)^3 \left(1 - e_{\rm out}^2\right)^{3/2}  \nonumber \\
&= \frac{n}{n_{\rm out}^2}  \left(1 - e_{\rm out}^2\right)^{3/2} 
\end{align} 
We ignore $e_{\rm out} \sim 0.04$.  We want to express \Eq{eq:doubleav} as a condition on 
\begin{align}
f_{\rm H}  &\equiv \frac{a}{R_{\rm H}}  = \frac{a}{a_{\rm out}} \left(\frac{3 m_3}{m_1+m_2}\right)^{1/3} = 3^{1/3} \left( \frac{n_{\rm out}}{n} \right)^{2/3}
\end{align} 
with $R_{\rm H}$ the inner binary Hill radius.  

The largest eccentricity of the inner orbit, $e$, is given by the collision condition at perihelion:
\begin{align}
1 - e &=b_{\rm c}/a \ll 1\, .
\end{align} 
 The collisional impact parameter $b_{\rm c}$ will depend in detail on the sizes and shapes of the two bodies.  We thus define an order unity radius ratio $f_D \equiv b_{\rm c}/\bar{D}$ where $(m_1+m_2) = (\pi /8) \rho_\bullet \bar{D}^3$ defines the effective diameter of the binary assuming equal densities.  For a large sphere and a much smaller body,   $f_D = 1/2$, and for two equal size spheres, $ f_{D} = 1/2^{1/3}$.  From \citet{Porter+19} for the dimensions of MU69 $f_{D} = 1$ for a collision along the long axis.   Using $1 + e \simeq 2$, we  calculate
\begin{align}
1 - e_{\rm max} ^2 &\simeq \frac{2 b_{\rm c}}{a} = \frac{2 f_D \bar{D}}{f_{\rm H} R_{\rm H}}  =\frac{ 4 f_D}{f_{\rm H}} \left(\frac{3}{ \pi} \right)^{1/3}  \left(\frac{n_{\rm out}^2}{G \rho_\bullet} \right)^{1/3} \, .
\end{align} 
We can  express (again ignoring $e_{\rm out}$):
\begin{align}
\frac{T_{\rm out}}{t_{\rm kozai}}  &= \frac{2 \pi f_{\rm H}^{3/2}}{\sqrt{3} }
\end{align} 
and thus from \Eq{eq:doubleav} the double averaged approximation should be valid for
\begin{align}
f_{\rm H} &\lesssim  \frac{3^{1/3}}{\pi^{7/12}} {f_D}^{1/4} \left(\frac{n_{\rm out}^2}{G \rho_\bullet}\right) ^{1/12} \nonumber\\
&\simeq 0.09 \left(\frac{0.5 \text{ g cm}^{-3}}{\rho_\bullet}\right)^{1/12} \left(\frac{44.5 \text{ AU}}{a_{\rm out} }\right)^{1/4}  
\end{align} 

So double-average secular dynamics should be applicable to
approximately $f_H \lesssim 0.1$. 

By comparison the single-averaged approximation is valid for 
\begin{align}\label{eq:singleav}
t_{\rm kozai} \sqrt{1 - e^2} & \gtrsim T
\end{align}
or in Hill units and again for $e_{\rm out} = 0$
\begin{align}
f_{\rm H} &\lesssim \left(\frac{3}{\pi} \right)^{1/3} f_D^{1/7}  \left(\frac{n_{\rm out}^2}{G \rho_\bullet}\right) ^{1/21} \nonumber\\
&\simeq 0.3 \left( \frac{0.5 \text{ g cm}^{-3}}{\rho_\bullet}\right)^{1/21} \left(\frac{44.5 \text{ AU}}{a_{\rm out} }\right)^{1/7}  
\end{align} 

We show in \fig{fig:SinglevsDouble} a comparison between the
double-averaged model and the single-averaged model for four values of
the Hill radius fraction: 0.04, 0.1, 0.2, and 0.4. The upper panels
show the eccentricity, and the lower panels the inclination. One full
Kozai-Lidov cycle is shown for each semimajor axis. The double-averaged
model is as presented in \eq{eq:model1} to \eq{eq:model4}, ignoring
the tides and dissipation terms. The single-averaged equations are \citep{Vashkovyak05,Shevchenko17}

\beqn
\frac{da}{dM_\beta} &=&0\\
\frac{de}{dM_\beta} &=& 10e\left(1-e^2\right)^{1/2}\left[\sinisq\sintwow + \left(2-\sinisq\right)\sintwow\costwop + 2\cosI\costwow\sintwop\right]\\
\frac{dI}{dM_\beta}  &=& -2\sinI \left(1-e^2\right)^{-1/2}\left\{5e^2\cosI \sintwow \left(1 -\costwop\right) - \left[2 + e^2\left(3 + 5\costwow\right)\right] \sintwop\right\}\\
\frac{d\omega}{dM_\beta}  &=& 2 \left (1-e^2\right)^{-1/2}   \left\{4 + e^2 - 5\sinisq +5 \left(\sinisq - e^2\right)\costwow + 5 \left(e^2 - 2\right)\cosI\sintwow\sintwop \right.\nonumber\\
&&\left.+ \left[5\left(2 - e^2 - \sinisq\right)\costwow - 2 - 3e^2 + 5\sinisq\right]\costwop\right\}\\
\frac{d\Psi}{dM_\beta}&=&  -\nu - 2\left(1-e^2\right)^{-1/2} \left\{ \left[2 + e^2\left(3 - 5\costwow\right)\right]\cosI\left(1 - \costwop\right) - 5e^2\sintwow\sintwop\right\}
\eeqn

\noindent where $\nu \equiv 16/3 \ (n/n_{\rm out})$, and the
quantity $M_\beta\equiv \beta M$ is a scaled mean anomaly where
$\beta \equiv 3/16 \  (n_{\rm out}/n)^2$. The quantity $\Psi$ is related to the longitude of the ascending
node via $\Psi \equiv \varOmega - \nu M_\beta$. 

We reproduce that the double-averaged model is applicable up to 0.1 $R_H$. Beyond this
radius the single-averaged model starts to show oscillations on top of
the double-average prediction, of increasing amplitude as we increase
the semimajor axis. These extra oscillations, reaching values of
eccentricity beyond the predicted by the double-averaged model, will
make contact more likely. Our solution based on the double-averaged
model is thus a conservative estimate of contact. Notice also that the
bound of the inclination oscillations also changes, allowing for
values lower than the original inclination, which is not possible in
the double-average model. As a result we cannot draw conclusions on
final inclination based on the double-averaged model.

\end{document}